\documentclass[12pt]{article}
\usepackage[sectionbib]{chapterbib}
\usepackage{latexsym,diagbox,graphicx,amssymb,amsmath,bm,subfigure,textcomp,wrapfig,url,enumitem,relsize,hyperref,makecell,multicol,float,lscape,amsmath,longtable,setspace,cite,times}
\usepackage[pdftex,dvipsnames]{xcolor}
\usepackage[margin=1in]{geometry}
\usepackage[font=small,labelfont=bf]{caption}

\newcommand{\R}{\mathbb{R}}

\newcommand{\bff}{{\bf f}}
\newcommand{\bfg}{{\bf g}}

\newcommand{\bfm}{{\bf m}}
\newcommand{\bfn}{{\bf n}}

\newcommand{\bfr}{{\bf r}}

\newcommand{\bfv}{{\bf v}}
\newcommand{\bfw}{{\bf w}}
\newcommand{\bfx}{{\bf x}}
\newcommand{\bfy}{{\bf y}}

\newcommand{\bfF}{{\bf F}}

\newcommand{\bfI}{{\bf I}}

\newcommand{\bfR}{{\bf R}}

\newcommand{\bfT}{{\bf T}}

\newcommand{\beq}{\begin{equation}}
\newcommand{\eeq}{\end{equation}}
\newcommand{\beqs}{\begin{eqnarray}}
\newcommand{\eeqs}{\end{eqnarray}}

\newcommand{\calI}{{\cal I}}

\newcommand{\calS}{{\cal S}}

\setcellgapes{10pt}
\hypersetup{colorlinks=true,linkcolor=black}
\begin{document}

\begin{center}
{\LARGE  \bf Comprehensive scaling laws across 

\vspace{5mm} animals, microorganisms and plants} 

\vspace{10mm}
Huan Liu\footnote{
Department of Aerospace Engineering and Mechanics, University of Minnesota, Minneapolis, MN 55455, USA}, Shashank Priya\footnote{
Department of Chemical Engineering and Materials Science, University of Minnesota, Minneapolis, MN 55455, USA}, and  Richard D. James$^1$

\vspace{8mm}
\end{center}
\normalsize

\noindent {\normalsize {\bf Abstract.} 
{\bf Scaling laws illuminate Nature's fundamental biological principles and guide bioinspired materials and structural designs.  In simple cases they are based on the fundamental principle that all laws of nature remain unchanged  (i.e., invariant) under a change of units.  A more general framework is a change of variables for the governing laws that takes all equations, boundary, and interaction conditions
into themselves. We consider an accepted macroscale system of partial differential equations including coupled fluid dynamics, nonlinear elasticity, and rigid body mechanics for a complex organism.  We show that there is a set of scaling laws
where length, time, density, elastic modulus, viscosity, and gravitational constant undergo
nontrivial scaling (Table 1).  We compare these results to extensive data sets mined from the literature on beating frequency of flying, swimming, and running animals, speed of bacteria, insects, fish, mammals and reptiles, leg stiffness of mammals, and modulus of elasticity of plants.  The uniform agreement of the scaling laws with the dynamics of fauna, flora, and microorganisms supports the
dominating role of coupled nonlinear elasticity and fluid dynamics in evolutionary development. We conclude with predictions for some prehistoric cases for which observations are unavailable.}
} 

\vspace{5mm}
Evolution drives performance and adaptation of a species, subject to a complex set of constraints imposed by a dynamic environment.  Among these constraints are fundamental physical laws, which, despite the complexity of Nature, may have universal implications -- Nature's design principles -- for broad classes of flora and fauna.  One family of such universal implications is
scaling laws.  Discovery of these scaling laws enables a better understanding of Nature's design principles and informs strategies for the creation of sustainable materials and structures.

There are various approaches to scaling  -- also termed in subcases similarity, dynamical similarity, or similitude -- appropriate for different forms of the underlying theory.  If
the theory is governed by a finite set of algebraic relations among physical quantities, then one can invoke the fundamental physical principle that all laws of physics must be invariant under a change of units. The resulting system of equations can be exploited using the Buckingham $\pi$ Theorem.

Most general physical theories are governed by partial differential equations.  These theories are
infinite-dimensional, and the Buckingham $\pi$ Theorem
is not applicable, but the principle of invariance
under a change of units holds.  This is frequently exploited by recasting the governing equations in dimensionless form.  

A more general approach adopted here, that subsumes non-dimensionalization, is to seek a transformation
of the governing partial differential equations, i.e., a change of variables,  which takes a
solution into another solution of the same equations
and preserves the interaction and boundary conditions.  A general change of variables is allowed: both dependent and independent variables can be transformed and the transformations need not be linear. The
interaction conditions include the equality of motions
and forces at places where solid parts meet each other or the fluid.     

A simple illustrative example of these distinctions concerns heat conduction  in a bar, governed by
the partial differential equation $\theta_t = k  \theta_{xx}$ known as the heat equation, where 
$\theta(x,t)$ is the temperature at position $x$ and time $t$, $k$ is the thermal diffusivity (units length$^2$/time)
and the subscripts denote partial derivatives.  Any solution $\theta(x,t)$ of the heat equation can be made dimensionless 
in the standard way by defining 
$\Theta(\xi, \tau) = (1/ c) \theta(\xi L, \tau T) $
with $c, L, T$ suitable constants with dimensions of temperature, length, and time, respectively. 
Substitution of $\theta(x,t) = c\, \Theta (x/L,t/T)$
into $\theta_t = k \theta_{xx}$ gives the equivalent non-dimensional equation:
$\Theta_{\tau} =  K \Theta_{\xi \xi}$, where
$K = kT/L^2$ is dimensionless.  But other more subtle changes of variables cannot be obtained by non-dimensionalization.  For example, suppose again
that $\theta(x,t)$ satisfies $\theta_t = k \theta_{xx}$ and define \cite{olver1993applications}
\beq
u(x,t) = e^{(1/k)(-\delta x + \delta^2 t)} \theta(x-2 \delta t, t).   \label{po}
\eeq
Then $u(x,t)$ also satisfies $u_t = k u_{xx}$. This cannot be obtained by non-dimensionalization, since
(\ref{po}) contains the new dimensional constant
$\delta$ (units length/time) that does not appear in the heat equation.  

Here, we are motivated by a famous video of a dead fish swimming upstream ``when its flexible body resonates with oncoming vortices''
\cite{beal_et_al_2006} in a complex flow. For there to be resonant interactions, the elasticity of the fish body must be relevant.  In this case we interpret elastic properties in this context: the elasticity of a recently deceased specimen. Live fish also utilize this feature to reduce energy consumption by altering their motions to match the environmental vortex shedding frequency \cite{liao2003fish,liao2003karman}. The elastic properties of plants have also been identified \cite{mcmahon1976tree} as relevant to their 
survival under wind and gravitational forces. 
To describe elastic properties, we use two nonlinear elastic laws considered to be exceptionally reliable: the Gent model \cite{gent1996new,horgan2015remarkable} for soft tissue and the St.~Venant-Kirchhoff model\cite{truesdell2004non} for stiff tissue.
Both of these models are accurate for large deformations and rotations, and
they are parameterized in terms of easily measured physical constants (Young's modulus, shear modulus, and Poisson's ratio).  This permits a variety of scaling laws
despite our general setting.

Motivated by these observations we derive in Methods rigorous scaling laws in a general setting from the coupled
partial differential equations of
nonlinear elasticity, rigid body mechanics, and fluid dynamics, including gravitational forces. The results are summarized in 
Table \ref{table1}. They allow for complex motion and turbulence and 
specialize to regimes appropriate to inviscid and Stokes flow.  To have a nontrivial scaling for general materials, we deduce that conditions of geometric similarity must hold.
Geometric similarity is consistent with observations on a  wide collection of birds, fish, and insects
in a size range from insects to cetaceans \cite{lucas2014bending, blickhan1993similarity}. Drawing on a broad collection of observations in
the literature, we find below remarkable agreement between
the derived scaling laws and measured properties
of microorganisms, insects, reptiles, amphibians, crustaceans, fish, birds, running, flying, and swimming mammals, and plants.

While few authors deduce scaling laws from a rigorous treatment of the full equations of motion (including coupled fluid dynamics and elasticity), semi-empirical laws have been widely reported.  The zoologist Alexander observed relationships between walking/running speed, stride length, and body size of animals \cite{alexander1976estimates}, and later he proposed a dynamic similarity hypothesis for the gaits of quadrupedal mammals \cite{alexander1983dynamic,alexander2003principles}, which agrees well with experimental data. Norberg and Rayner \cite{norberg1987ecological} found that a regression model predicts with high precision the behavior of bats whose dimensions are geometrically similar.   
Heglund et al.~\cite{heglund1974scaling} concluded that the stride frequency of animals from mice to horses changes with size in a predictable manner. Taylor et al.~ \cite{taylor2003flying} found that the Strouhal number lies in a narrow range when flying and swimming animals cruise, which is also implied by the finding that the forward velocity of large fish is proportional to the product of tail beat frequency and tail beat amplitude \cite{gazzola2014scaling}.



\begin{table}
\caption{\textbf{Scaling laws.} Scaling laws with scaling parameters $\lambda$ (geometry), $\alpha$ (time), and $\eta$ (density).}\label{table1}
\vspace{3mm}
\renewcommand{\arraystretch}{2}
\setlength\tabcolsep{1.5pt}
\setlength\extrarowheight{-3pt}
\centering
\begin{tabular}{l|cc|cc|cc|cc}
\hline
    & \multicolumn{2}{c|}{Rigid parts}    & \multicolumn{2}{c|}{Stiff tissue}   & \multicolumn{2}{c|}{Soft tissue}    & \multicolumn{2}{c}{Fluid}   \\ \cline{2-9}
Variable  & \makecell[c]{\footnotesize Organism \\ 1 }  & \makecell[c]{\footnotesize Organism \\ 2} & \makecell[c]{\footnotesize Organism \\ 1}   & \makecell[c]{\footnotesize Organism \\ 2}   & \makecell[c]{\footnotesize Organism \\ 1}   & \makecell[c]{\footnotesize Organism \\ 2}   & \makecell[c]{\footnotesize Organism \\ 1}   & \makecell[c]{\footnotesize Organism \\ 2}   \\
\hline
\multicolumn{9}{c}{Scaling laws of geometry, time, and material properties}    \\ \cline{1-9}
Geometry                       & $\bfx$  & $\lambda\bfx$   & $\bfx$  & $\lambda\bfx$ & $\bfx$  & $\lambda\bfx$               &$\bfr$  & $\lambda\bfr$ \\
Time      
& $T$         & $\alpha T$           
& $T$         & $\alpha T$ 
& $T$         & $\alpha T$ & $T$         & $\alpha T$  \\ 
Density       
& $\rho_{r}$         & $\eta\rho_{r}$           
& $\rho_{st}$         & ${\eta}\rho_{st}$                
& $\rho_{so}$         & ${\eta}\rho_{so}$& $\rho_f$         & ${\eta}\rho_f$ \\
Young's modulus
&          & 
& $E$         & $\frac{\eta\lambda^2}{\alpha^2} E$        &&      &          &  \\
Poisson's ratio
&          & 
& $\hat\nu$         & $\hat\nu$               &         &   &        &  \\
Shear modulus
&          & 
&         &              &    $\mu$&$\frac{\eta\lambda^2}{\alpha^2}\mu$&      &  \\
Bulk modulus
&          & 
&         &              &    $\kappa$&$\frac{\eta\lambda^2}{\alpha^2}\kappa$&      &  \\
Kinematic viscosity
&         &             &         &      & & &   $\nu$       &  ${\frac{\lambda^2}{\alpha}}\nu$\\
Gravity
&    $\bfg$     &         $\frac{\lambda}{\alpha^2}$ $\bfg$  &    $\bfg$     &   $\frac{\lambda}{\alpha^2}\bfg$    &   $\bfg$       & $\frac{\lambda}{\alpha^2}$ $\bfg$ &   $\bfg$       & $\frac{\lambda}{\alpha^2}\bfg$  \\
\cline{1-9}
\multicolumn{9}{c}{Scaling laws of motions and forces}    \\ \cline{1-9}
Motion  
& $\bfy_r$         & $\lambda\, \bfy_r$          & $\bfy_{st}$         & $\lambda\, \bfy_{st}$     & $\bfy_{so}$         & $\lambda\, \bfy_{so}$        
&& \\
Velocity
& $\bfv_r$         & ${\frac{\lambda}{\alpha}}\bfv_r$ &       $\bfv_{st}$         & ${\frac{\lambda}{\alpha}}\bfv_{st}$  &$\bfv_{so}$         & ${\frac{\lambda}{\alpha}}\bfv_{so}$    & $\bfv_f$         & ${\frac{\lambda}{\alpha}}\bfv_f$  \\
Angular velocity   & $\omega_r$         & $\frac{1}{\alpha}\omega_r$     &       $\omega_{st}$         & $\frac{1}{\alpha}\omega_{st}$  &$\omega_{so}$         & $\frac{1}{\alpha}\omega_{so}$    &          &  \\
Frequency   & $f_r$         & $\frac{1}{\alpha}f_r$     &       $f_{st}$         & $\frac{1}{\alpha}f_{st}$  &$f_{so}$         & $\frac{1}{\alpha}f_{so}$    &          &  \\
Pressure
&      &    &       &  &   $P_{so}$       &  $\frac{\eta\lambda^2}{\alpha^2} P_{so}$               &   $P_f$       &  $\eta\frac{\lambda^2}{\alpha^2} P_f$\\
Force
 &   $\bff_r$       &  $\frac{\eta\lambda^4}{\alpha^2} \bff_r$       &   $\bff_{st}$       &  $\frac{\eta\lambda^4}{\alpha^2} \bff_{st}$      &   $\bff_{so}$       &  $\frac{\eta\lambda^4}{\alpha^2} \bff_{so}$ &        &  \\
Moment
 &   $\bfm_r$       &  $\frac{\eta\lambda^5}{\alpha^2} \bfm_r$       &   $\bfm_{st}$       &  $\frac{\eta\lambda^5}{\alpha^2} \bfm_{st}$      &   $\bfm_{so}$       &  $\frac{\eta\lambda^5}{\alpha^2} \bfm_{so}$ &        &  \\
\hline
\multicolumn{9}{c}{Scaling laws of other associated variables}    \\ \cline{1-9}
Mass  
& $M_r$         & $\eta\lambda^3 M_r$ 
& $M_{st}$         & $\eta\lambda^3 M_{st}$ & $M_{so}$         & $\eta\lambda^3 M_{so}$              &          &  \\
Inertia 
& $\calI_r$         & $\eta\lambda^5 \calI_r$ 
&        & &         &               &          &  \\
Axial stiffness
 &        &      &   $K_{st}$       &  $\frac{\eta\lambda^3}{\alpha^2} K_{st}$      &   $K_{so}$       &  $\frac{\eta\lambda^3}{\alpha^2} K_{so}$ &        &  \\
Deformation gradient   & $\bfR$         & $\bfR$            & $\bfF_{st}$         & $\bfF_{st}$  & $\bfF_{so}$         & $\bfF_{so}$        &          &  \\
Cauchy Stress
&         &        &$\boldsymbol \sigma_{st}$        &    $\frac{\eta\lambda^2}{\alpha^2}\boldsymbol \sigma_{st}$       & $\boldsymbol \sigma_{so}$  &  $\frac{\eta\lambda^2}{\alpha^2}\boldsymbol \sigma_{so}$& $\boldsymbol \sigma_f$  &  $\frac{\eta\lambda^2}{\alpha^2}\boldsymbol \sigma_f$\\
\hline
\end{tabular}
\end{table}

\begin{figure}
\centering
\includegraphics[width=\textwidth]{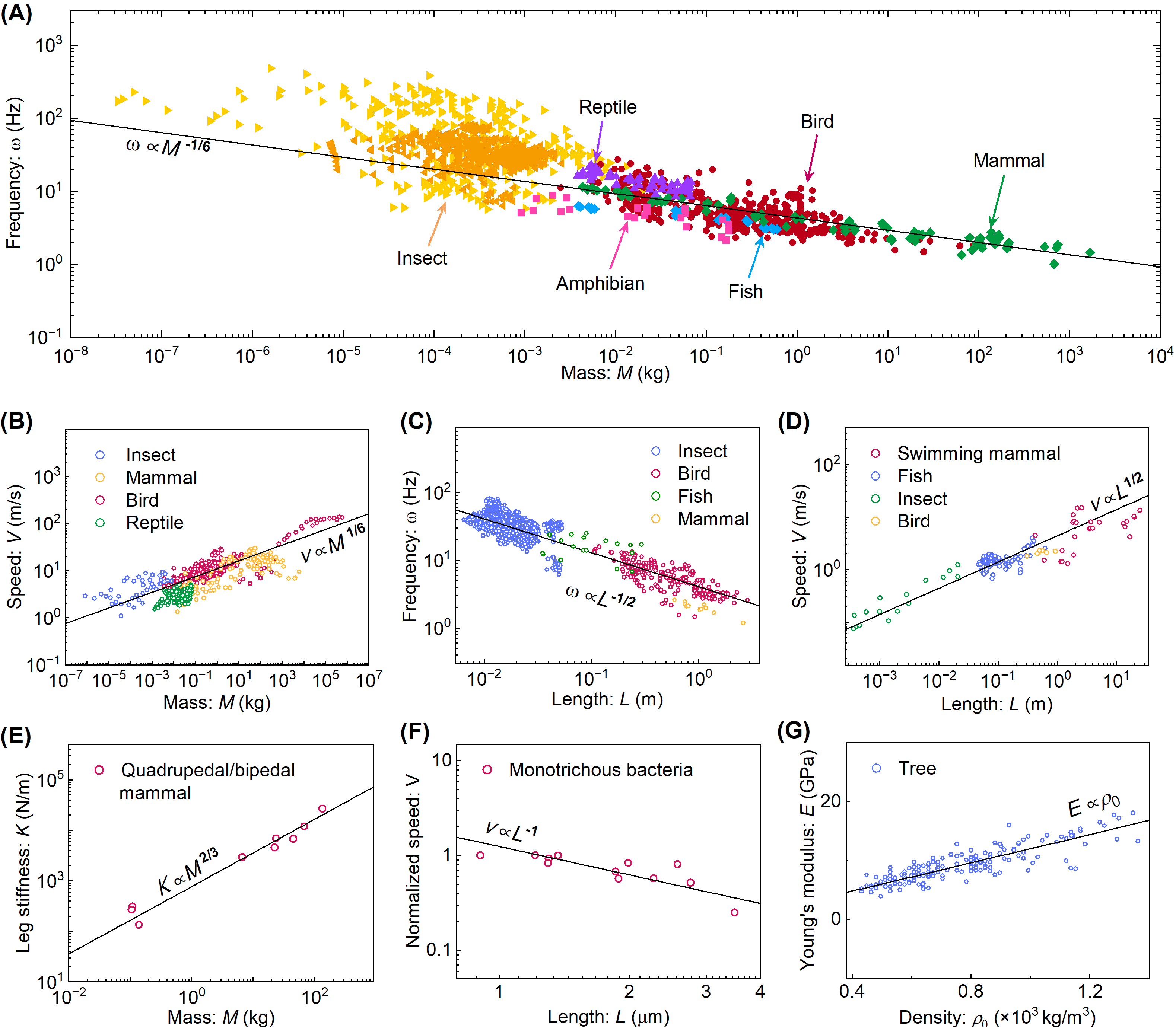}
\caption{\small Comparison between experimental data (dots) and the predictions (solid lines) by scaling laws. (A) Frequency-mass relation of flying, swimming, running, and crawling animals. (B) The scaling law between speed and body mass for animals. (C) The frequency-length relation of animals. For fish, the frequency refers to tail-beat frequency and length refers to the body length; for birds and insects, the frequency refers to wing-beat frequency and length refers to the wing span; for mammals, the frequency refers to the stride frequency and length refers to leg length.  (D) The speed-length relation of animals.  (E) The scaling law between leg stiffness and body mass.   (F) The scaling law between velocity and the body length of microorganisms, where the speed $V$ is normalized by the speed $V_0$ of the smallest observed bacteria, i.e., $V=$ real speed / $V_0$.  (G) The scaling law of Young's modulus to the density of trees. A more detailed discussion of the references used here and additional data are found in the Database in the Supplement Materials. }
\label{fig:collection}
\end{figure}

Microorganisms like bacteria often show certain speed-size correlations \cite{li2010experimental}.  In some studies  E. coli show subtle speed-size relations that disagree with our scaling laws.  However, as pointed out by Kamdar et al.~\cite{kamdar2023multiflagellarity} such cases
involve varying numbers of flagella during growth, which breaks geometric similarity.  In the case of E. coli we carefully choose data
consistent with geometric similarity (Figure \ref{fig:collection}(F), with details in the Database).

Following the pattern of argument in the analysis of a vertical axis wind turbine \cite{liu2023options}, we consider general changes of variables of the widely accepted dynamic theory of nonlinear elasticity coupled via boundary conditions to a Navier-Stokes fluid and rigid bodies. Our analysis (Methods) is straightforward and rigorous but slightly technical because nonlinear elasticity is formulated in Lagrangian form while the Navier-Stokes fluid is formulated in Eulerian coordinates.

Nevertheless, we
find a single nontrivial set of scaling laws (Table \ref{table1}) involving geometry, time, density, mass, Young's modulus, shear modulus, Poisson's ratio, viscosity,
gravitational constant, motion, velocity, angular velocity, frequency, pressure, force, moment, mass, inertia, and 
stiffness.  We then scoured the literature and found abundant data sets that are consistent with geometric similarity and therefore can be compared directly to our results. These references were used primarily for the data they present; the data are collected and curated in the Database. We did
not selectively choose data (``cherry picking'') from these sets, but we present all data that apply.  The results
demonstrate remarkable consistency between the scaling laws
(Table \ref{table1}) and experimental measurements in fish, birds, reptiles, amphibians, crustaceans, flying/swimming/running mammals, insects, microbes, and trees (Figure \ref{fig:collection}). 

In the following paragraphs we describe some special cases that are summarized by the diagrams in Figure \ref{fig:collection}.  To make the results widely interpretable, we present the relevant cases of Table 
\ref{table1} as an equality between dimensionless numbers
when this is possible.  To do this, we have to introduce
an apparently unrecognized dimensionless number
$Em = \rho V^2/E$, where $\rho$ is the density of the elastic tissue, $V>0$ is the characteristic velocity (of the fish, bird, reptile, etc.) and $E$ is Young's modulus for stiff elastic tissue and shear modulus for soft elastic tissue. $Em$
can be viewed as the square of the ratio of the body's velocity to its longitudinal elastic wave speed (stiff elastic tissue) or shear wave speed (soft elastic tissue). Thus, $Em$ is an analog for
elastic bodies of the Mach number except that both the
wave speed and the velocity refer to the elastic tissue. 
Also used below are Reynolds ($Re = V L/\nu$), Froude ($Fr = V/\sqrt{gL}$), and Strouhal ($St=f L/V$) numbers. Throughout the paper superscript $(s)$ represents small creatures and $(\ell)$ represents large creatures.

\paragraph{Inertial regime: fish, birds, reptiles, amphibians, crustaceans, and mammals} It is well accepted that the swimming behavior of fish larvae (typically $<5\ mm$) is dominated by the viscosity, while the swimming of fish adults (typically $Re>2000$; in many cases ${\sim}10^5$) is strongly determined by inertial forces \cite{muller2000hydrodynamics,voesenek2018biomechanics},
as is also consistent with Lighthill's elongated body theory \cite{lighthill1960note,lighthill1971large}.
Precisely the same derivation of scaling laws presented in Methods
is valid for this inertial regime, with the Navier-Stokes
equations replaced by the (inviscid) Euler equations.  The
effect of this change on Table \ref{table1} is simply to
put $\nu = 0$, which implies that $\lambda^2/\alpha$ is
not restricted by viscosity scaling. Assume smaller fish and larger fish live in the same fluid environment with the same fluid density and gravity. We have from Table 1,
\beq
Fr^{(s)}=Fr^{(\ell)},\quad Em^{(s)}=Em^{(\ell)},\quad St^{(s)}=St^{(\ell)}.\label{fluid_sea}
\eeq
We use $L$, $f$, $M$, $v$, $E$, and $K$ to represent the characterized length, frequency, mass, speed, elastic modulus, and stiffness of large $(\ell)$ fish  and corresponding values with subscript $(s)$ for the
small fish. Then we have $\lambda=L^{(\ell)}/L^{(s)}$ and, directly from Table 1,
\beq
\begin{aligned}
\frac{f^{(\ell)}}{f^{(s)}} =  \left[\frac{L^{(\ell)}}{L^{(s)}}\right]^{-\frac{1}{2}}= \left[\frac{M^{(\ell)}}{M^{(s)}} \right]^{-\frac{1}{6}},&\quad
\frac{v^{(\ell)}}{v^{(s)}} = \left[\frac{L^{(\ell)}}{L^{(s)}}\right]^{\frac{1}{2}} =   \left[\frac{M^{(\ell)}}{M^{(s)}}\right] ^{\frac{1}{6}},   \\
\frac{E^{(\ell)}}{E^{(s)}} \ = \ \frac{L^{(\ell)}}{L^{(s)}}\  =  \ \left[\frac{M^{(\ell)}}{M^{(s)}}\right]^{\frac{1
}{3}}, &\quad
\frac{K^{(\ell)}}{K^{(s)}} = \left[\frac{L^{(\ell)}}{L^{(s)}}\right]^2 =  \left[\frac{M^{(\ell)}}{M^{(s)}}\right]^\frac{2}{3}.
\end{aligned}
\label{scale_fish}
\eeq
(See {\bf Example of the use of Table 1} in Methods.)  Note the unusual fractional power scalings.
As the Reynolds number of most birds and swimming mammals is of order $10^5$, we include them.
These results are consistent with diverse 
measurements on fish, birds, reptiles, amphibians, crustaceans, and mammals. Relevant data are reproduced in Figure \ref{fig:collection}(A-E)
together with the predicted scaling laws (\ref{scale_fish}).
Note that Figure \ref{fig:collection}(A) covers a range of mass ratio of $10^{10}$.

Reynolds numbers of flying/hopping insects range from $10$ to $10^4$, which lies in between
the two flow regimes\cite{wang2005dissecting}: 1) Stokes flow around swimming bacteria, and 2) inviscid flows around birds,
fish,  and also running 
reptiles, amphibians, crustaceans, and mammals. In this intermediate regime, Table 1 gives only trivial scaling relations. Nevertheless, it can be seen that there is some level
of agreement of the scaling law of Figure \ref{fig:collection}(A) for insects (brown and yellow points). The yellow points are summarized by
Dudley\cite{dudley2002biomechanics}, which includes a wide variety of species of insects that do not follow geometric similarity. More generally, nontrivial geometric
similarity over a wide range is rare in most
insects, e.g., butterflies typically
stop growing after they change into winged 
adults.

\paragraph{Viscous regime: microorganisms}
As the Reynolds number decreases to near $0$, the viscous force plays an increasingly important role.  This is the case for microorganisms \cite{lauga2009hydrodynamics,lighthill1976flagellar}. Then the Navier-Stokes equations are simplified to Stokes equations \cite{lauga2009hydrodynamics}
but our scaling analysis and Table 1 are unchanged.  According to the scaling laws for Stokes flow and using Gent's model for a soft elastic tissue, we have 
\beq
Re^{(s)}=Re^{(\ell)},\quad Em^{(s)}=Em^{(\ell)},\quad St^{(s)}=St^{(\ell)}.\label{micro}
\eeq
Then the scaling laws of velocity and frequency are again found in Table 1:
\beq
\frac{v^{(\ell)}}{v^{(s)}} = \left[\frac{L^{(\ell)}}{L^{(s)}}\right]^{-1} =   \left[\frac{M^{(\ell)}}{M^{(s)}}\right]^{-\frac{1}{3}},\quad \frac{f^{(\ell)}}{f^{(s)}} = \frac{E^{(\ell)}}{E^{(s)}} =  \left[ \frac{L^{(\ell)}}{L^{(s)}}\right]^{-2} =  \left[\frac{M^{(\ell)}}{M^{(s)}} \right]^{-2/3}.
\eeq
See Figure \ref{fig:collection}(F).

\paragraph{Scaling laws in plants}
The scaling laws of Table 1 are also consistent with the elastic properties of plants. Mature trees show a nearly constant ratio of modulus of elasticity to density ($E/\rho_{st}$) \cite{niklas1993influence,mcmahon1973size,ngadianto2020wood}. We use the stiff model and consider various geometrically similar trees in a common location subject to typical maximum wind speed $V$.  Table 1 gives
\beq
Em^{(s)}=Em^{(\ell)},
\eeq
so $\frac{E}{\rho_{st}}\propto V^2\approx\, const.$, which implies
\beq
\frac{E^{(\ell)}}{E^{(s)}}=\frac{\rho^{(\ell)}_{st}}{\rho^{(s)}_{st}}\ .\label{plants}
\eeq
See Figure \ref{fig:collection}(G). 
Note that for this case $\frac{1}{Em}{\sim} 10^6$, while the two numbers $\frac{1}{Re}<1$ and $\frac{1}{Fr^2}<1$. 
The law (\ref{plants}) is also obeyed by large families of angiosperms (altogether, ${\sim}$300,000 species). Based on a regression model \cite{fournier2006tree}, the modulus of elasticity of the green wood of 
a wide range of angiosperm species  satisfies $E=10400(\frac{\rho_{st}}{0.53})^{1.03}$, which agrees well with (\ref{plants}).

\paragraph{Prehistoric fish, birds and apes}

This broad agreement with observed behavior suggests that these laws could be applicable to cases where dynamical measurements are impossible but, based on the fossil record, a geometrically similar surrogate can be found among present-day flora and fauna. 

The iconic giant mackerel shark, {\bf Otodus megalodon}, 
is known from fossilized teeth and vertebrae. These fossilized bones are found to be approximately geometrically similar to those of a great white shark, but 3$\sim$4 times larger in linear dimensions \cite{megalodon}. Great white sharks with a typical length of 4.5 $m$ and a weight of 900 $kg$ can swim at 7 $m/s$. Assuming geometric similarity, our scaling laws  (\ref{scale_fish}) imply that the body size, weight, and speed of a megalodon could reach $18$ $m$, 58 tons, and 14 $m/s$, respectively. In addition, the tail-beat frequency of great white sharks is about $0.4$ Hz on average, which according to our scaling laws implies megalodon had a tail-beat frequency of about $0.2$ Hz. 
 
 {\bf Pelagorni} is a genus of prehistoric birds
 having exceptionally large wingspan, thought to be close relatives of storks \cite{mayr2009paleogene}.  A conservative estimate of the
 wingspan of Pelagornis sandersi is  $6.4\ m$ \cite{ksepka2016giants,ksepka2014flight}. The current day marabou stork (Leptoptilos crumenifer) has a wingspan of $3.2\ m$, a mass of $5\ kg$, and a flight speed of around $45\ km/h$.  It flaps with a frequency of $2.4$ Hz in level flight \cite{kahl1966comparative}. It follows from the scaling laws (Table 1) that Pelagornis sandersi would have had a mass of around $40\ kg$, a speed up to $63.6\ km/h$, and a beating frequency of $1.7$ Hz. Our estimated speed is consistent with the inference from a sophisticated computer model, Flight-1.25 \cite{ksepka2014flight}. 

The apes of the genus {\bf Gigantopithecus} (Early to Mid-Pleistocene) are thought to be the largest primates that ever lived. From studies of fossilized teeth and jaw bones the body length of the Gigantopithecus blacki is estimated to be up to 3.7 $m$ \cite{ciochon1990other,zhang2017gigantopithecus}. Today's gorillas with a length of $1.7\ m$ can run at speeds of about  $32\ km/h$ \cite{iriarte2002differential}. Our scaling laws then imply that Gigantopithecus blacki could achieve speeds of 
a remarkable $47\ km/h$.  In addition, according to (\ref{fluid_sea}) and Table \ref{table1}, the force scales the same as volume: $\bff\propto L^3$. On average, a gorilla punch produces around $2,000$ pounds of force ($\approx 8.9\ kN$).  Then,  according to our scaling laws, the punching force of Gigantopithecus blacki could reach $20,620$ pounds of force ($\approx 91.7\ kN$), approximately 140 times the human punch force.

Not surprisingly, popular media give values that exceed some of these predictions.  For example, in the film The Meg, the length of the megalodon is about 23 $m$ \cite{themeglength} with a speed of about 41 $m/s$ \cite{themeg}, much higher than $15.8\ m/s$ from our scaling laws. In the 2005 movie, King Kong has a height of 7.6 $m$ \cite{kingkongheight} and a speed of about 40 $m/s$ (see Database), while our scaling laws 
predict a limit speed of around 18.8 $m/s$.

The understanding provided by Figure \ref{fig:collection} shows the commonality across diverse species for the utilization of a design principle that governs material and structural metrics. This principle paves the way for realizing quantitative bio-inspired performance.

\section*{Acknowledgments}

\paragraph*{Funding:}
All authors
acknowledge the support of the ONR MURI Grant N00014-23-1-2754. The work of RDJ and HL also benefited from a Vannevar Bush Faculty Fellowship (N00014-19-1-2623) and   AFOSR (FA9550-23-1-0093).
\paragraph*{Author contributions:}
R.D.J. and H.L. conceived the
project, conducted the research, and wrote the paper.
H.L. collected the data.  S.P. secured funding for the
project. All authors discussed the results, and reviewed and edited the manuscript.  
\paragraph*{Competing interests:}
There are no competing interests to declare.
\paragraph*{Data and materials availability:}
The Methods are included in the Supplementary Materials. All data used to compare with the scaling laws are referenced in the paper. References
primarily used for measured data are collected and curated under the Database in Supplementary materials.

\newpage
\appendix
\begin{center}
{\LARGE  \bf Supplementary Materials} 
\end{center}
\vspace{3mm}

\noindent{\LARGE  \bf Methods}

\vspace{5mm}
\noindent{\large \bf Structure of the derivation of scaling laws}
\vspace{2mm}

\noindent The scaling laws summarized in Table 
\ref{table1} generalize laws derived for the analysis of
a vertical axis wind turbine\cite{liu2023options} by including additional relations arising from gravitational forces and the Gent or St. Venant-Kirchhoff models of nonlinear elasticity (see below). As noted at the beginning of this article, the structure of the calculation begins with the Navier-Stokes equations of fluid dynamics
coupled via two-way interaction conditions to an elastic 
solid and a rigid body which themselves are coupled
to each other by interaction conditions (see below).  These are
written for the ``Organism 1'' (Table \ref{table1}) with given
dimensions and properties.  ``Organism 2'' 
refers to another species whose properties are
related in a certain way to  Organism 1 as described in Table \ref{table1}.  The analysis below shows that, if 
all equations, interaction conditions, and boundary conditions are satisfied
for the Organism 1 under some given conditions, then they are necessarily satisfied for Organism 2 under other (predicted) conditions. 
The results are summarized in 
Table \ref{table1}.

In applications considered here the fluid is typically water or air, 
the elastic parts include connective tissue (e.g., tendons, feathers), and the rigid parts are bone or shell. In plants, the stem or woody material is considered elastic
and the fluid part refers to the surrounding air. 

\vspace{5mm}
\noindent{\large \bf Kinematics: the Eulerian and 
Lagrangian descriptions of motion}
\vspace{2mm}

\noindent The Navier-Stokes equations are conventionally presented 
using a different kinematics
(Eulerian) than the equations of nonlinear elasticity
(Lagrangian).  Rigid body mechanics commonly uses
both forms, but we choose Lagrangian here.  To derive scaling laws we have to go back and forth
between these descriptions. The Lagrangian description
uses a reference domain $\Omega$ for the elastic parts and $\Omega_r$ for the rigid parts.  They are
usually taken as the shape occupied by the organism at $t = 0$.  Letting $\bfx \in \Omega$ (or $\Omega_r$), the motion
of the organism is described by a function $\bfy(\bfx,t)$ (resp.~$\bfy_r(\bfx,t)$ for rigid parts), giving the position of the ``particle''
$\mathbf x$ at time $t$, while the Eulerian description
assigns a velocity $\bfv(\bfy,t)$ to points $\bfy$ in space at the present time $t$.  The relation between the two descriptions is:
\begin{enumerate}
\item Given $\bfv(\bfy,t)$, find $\bfy(\bfx,t)$ by solving the ordinary differential equation $\dot \bfy (\bfx, t) = \bfv(\bfy(\bfx, t), t)$ with initial condition $\bfy(\bfx,0) = \bfx \in \Omega$.  The
superimposed dot denotes the time derivative.
\item  Given $\bfy(\bfx,t)$, find $\bfv(\bfy,t)$
by using the formula $\bfv(\bfy,t) = \dot{\bfy} (\bfy^{-1}(\bfy,t),t)$.  Here the inverse is
taken with $t$ fixed: $\bfy^{-1}(\bfy(\bfx, t),t) = \bfx$.
\end{enumerate}
These relations are used below, especially for the interaction conditions. In the following analysis, the quantities with subscripts $f$, $r$, $st$ and $so$ are for the fluid, the rigid body, the stiff elastic solid, and the soft elastic solid, respectively. Without causing ambiguity, we will omit the subscripts in some equations below for clarity.

\vspace{5mm}
\noindent{\large \bf The Navier-Stokes, St.~Venant-Kirchhoff and Gent models}
\vspace{2mm}

\noindent 
\begin{enumerate}[leftmargin=*]
    \item {\bf Navier-Stokes fluid}
The equations of motion of an incompressible Navier-Stokes fluid are given by
\beq
\bfv_t + \nabla_{\bfr} \bfv\, \bfv = -\nabla_{\bfr} p + \nu \Delta \bfv+\bfg, \quad {\rm div}\, \bfv = 0,   \label{ns2s}
\eeq
where $\bfr\in \R^3\setminus \big(\bfy(\Omega,t)\cup \bfy(\Omega_s,t) \cup \bfy_r(\Omega_r,t)\big).$ The constitutive equation for the fluid is
\beq
\boldsymbol \sigma_f = \rho_f\left(-p\bfI + \nu (\nabla \bfv + \nabla \bfv^T)\right).  \label{sig}
\eeq

\item {\bf Elastic body}
The equations of motion of the nonlinear elastic body 
\beq
\rho(\bfx) \ddot{\bfy} = {\rm div} _{\bfx}\bfT+\rho(\bfx)\bfg, \quad \bfx \in \Omega, \quad t>0, \label{blm1s}
\eeq
where $\rho$ is either $\rho_{st}$
(stiff material) or $\rho_{so}$ (soft material), see just below.
This is to be solved for $\bfy(\bfx,t), \ \bfx \in \Omega$.  The reference density $\rho(\bfx)$ is given on $\Omega$. The Cauchy stress is related to $\bfT$ by $\boldsymbol \sigma = \frac{1}{\det \bfF} \bfT \bfF^T$,where $\bfF = \nabla_{\bfx} \bfy$. There are several well-known constitutive equations for elastic solids. We use the following models in our analysis.

\vspace{2mm}
1) {\bf Stiff elastic materials}
We use the St.~Venant-Kirchhoff model to characterize stiff elastic tissue. The
corresponding (first) Piola-Kirchhoff stress is
\beqs
\bfT (\bfF) &=&  \frac{E}{2} \bigg(  \frac{ \hat{\nu}}{(1+ \hat{\nu})(1-2 \hat{\nu})}\, \big( {\rm tr}(\bfF^T \bfF) - 3) \big) \bfI  +  \frac{1}{1 + \hat{\nu}}\, \big(\bfF \bfF^T  - \bfI\big)    \bigg) \bfF,
\label{StVK}
\eeqs
where we have traded Young's modulus $E$ and Poisson's ratio $\hat{\nu}$ for
the Lam\'e moduli. The corresponding free energy is given by Raoult \cite{Raoult1986}. This model is highly accurate for deformation gradients near SO(3) (i.e., near rigid rotations) because it is exactly frame-indifferent
and agrees with linear elasticity near
$\bfF = \bfI$, but it is known to have unacceptable behavior far from SO(3).
Thus, it is accurate for stiff materials under modest stress that produces deformations near SO(3). The St.~Venant-Kirchhoff model has been used for biomechanical modeling 
of muscle and tendon in human diaphragm and thorax \cite{picinbono2001nonlinear}, animal organs \cite{ladjal2015biomechanical}, and facial tissue \cite{gladilin2003biomechanical}. 

\vspace{2mm}
2) {\bf Soft elastic materials}
We use Gent's phenomenological
constitutive model for hyperelastic incompressible soft tissue \cite{gent1996new,horgan2015remarkable}. The strain energy density function is represented as
\beq
\varphi^G=-\frac{\mu}{2} J_m 
\ln\left(1-\frac{I_1-3}{J_m}\right),\quad I_1<J_m+3,\ J_m\gg1,\label{gent_incomp2}
\eeq
where $\mu$ is the shear modulus, $I_1={\rm tr}(\bfF^{\rm T}\bfF)=\lambda_1^2+\lambda_2^2+\lambda_3^2$ is the first strain invariant, $J_m$ 
is a maximum permitted value of $J_1=I_1-3$, which  accounts for limited polymeric chain extensibility, and incompressibility means $\lambda_1\lambda_2\lambda_3=1$. The Cauchy stress  has the form 
\beq
\boldsymbol\sigma_{so}=-P_s\bfI+\frac{\mu J_m}{J_m-I_1+3}\bfF\bfF^{\rm T},\label{gent}
\eeq
As $J_m$ tends to infinity, (\ref{gent_incomp2}) reduces to the
classical neo-Hookean model: $\varphi^{nH}=\frac{\mu}{2} \left(I_1-3\right)$.  We assume $J_m$ varies little between materials. 

A widely  used compressible version is the compressible neo-Hookean model \cite{ogden1997non}, which is simple and typically accurate for strains less than 20$\%$. The strain energy density function is given by
\beq
\varphi^{nH}=C_1(I_1-3-2\ln J)+D_1(J-1)^2,
\eeq
where $C_1,D_1$ are material constants and $J={\rm det}(\bfF)=\lambda_1\lambda_2\lambda_3$, and the second term represents the pressure-volume response. At low strains, $C_1=\frac{\mu}{2},D_1=\frac{\kappa}{2}-\frac{\mu}{3}$ to be consistent with linear elasticity, where $\kappa$ is the bulk modulus. Then the Cauchy stress is found by
\beq
\boldsymbol\sigma_{so}=\mu\frac{1}{J}(\bfF\bfF^{\rm T}-\bfI)+(\kappa-\frac{2}{3}\mu)(J-1)\bfI.\label{comp_rubb}
\eeq
There are some other models for slightly compressible materials with $\mu\ll\kappa$, such as Blatz model \cite{blatz1962application}, Christensen model \cite{christensen1988two}, Levinson and Burgess model \cite{levinson1971comparison,burgess1972instability}. In these models, the Cauchy stress also has the form of $\boldsymbol\sigma_{so}=\mu F_1(I_1,I_2,I_3)+\kappa F_2(J)$, where $F_1,F_2$ are two functions of strain invariants.  Thus, our analysis applies to 
these compressible cases if the ratio $\kappa/\mu$ does not vary significantly among the species considered. The Neo-Hookean, Demiray, and one-term Ogden models 
used in biomechanical studies of soft tissue
\cite{budday2017mechanical} also are linear in the shear modulus $\mu$.
\end{enumerate}

\vspace{5mm}
\noindent{\large \bf Dimensionless numbers }
\vspace{2mm}

\noindent Some of our results in Table \ref{table1} can be seen by direct
non-dimensionalization of the equations of fluid
dynamics, elasticity, and rigid body mechanics. Bear in
mind that these are only necessary conditions for
dynamical similarity; in addition, the
interaction conditions have to be imposed as
treated below. For clarity we omit the subscripts $st$ and $so$ for elastic solids in the following analysis. The equations of motion for the fluid, solid elastic bodies, and rigid bodies have the forms of
\beq
\renewcommand{\arraystretch}{1.3}
\left\{\begin{array}{rcll}
\bfv_t + \nabla\bfv\, \bfv &=& -\frac{1}{\rho_f}\nabla P + \nu\Delta \bfv+\bfg,\quad {\rm div}\, \bfv = 0,&{\rm for\ fluids}, \\
\ddot{\bfy} &=&\frac{E}{\rho}{\rm div} _{\bfx}\bar\bfT+\bfg,&{\rm for\ elastic\ solids}, \\
M_r \ddot{\bfy}_{cm} &=&  \bff_r, \quad {\rm and} \\
\calI_r \dot{\bfw} &=& -\bfw \times \calI_r \bfw - \bfR^T\big(\bfy_{cm} \times \bff_r \big)+ \bfR^T\bfm_r, &{\rm for\ rigid\ bodies}
\end{array}\right.\label{eq_motion}
\eeq
where $\rho_f$ and $\rho$ are the density of the fluid and the density in the reference domain of the elastic body, respectively; $E$ represents the elastic modulus, which is Young's modulus for the stiff elastic body and shear modulus for soft elastic materials; $\bar\bfT$ is the nondimensionalized (first) Piola-Kirchhoff stress $\bfT$ satisfying $\bfT=E\bar\bfT$, where $E$ represents Young's modulus for the stiff elastic solid and the shear modulus for soft elastic solid. By substituting (\ref{fs}) and (\ref{ms}), the last of (\ref{eq_motion}) can be simplified to
\beqs
\calI_r \dot{\bfw} &=& -\bfw \times \calI_r \bfw + \int_{\partial\Omega_r\setminus\calS}(\bfx-\bfx_{cm})\times\sigma_f\bfR\bfn_0da_0+\int_{\calS}(\bfx-\bfx_{cm})\times\bfT\bfn_0da_0\nonumber\\
&=&-\bfw \times \calI_r \bfw +\bar\bfm_1+\bar\bfm_2,
\eeqs
where $\bar\bfm_1$ and $\bar\bfm_2$ correspond to the last two integrals, respectively. Let  $V>0,f>0$ and $L>0$ be the characteristic velocity, frequency, and length of Organism 1, and define
\beqs
&&\bfv^*=\frac{\bfv}{V},\; t^*=\frac{t}{L/V},\;P^*=\frac{P}{\rho_f V^2},\;\bfx^*=\frac{\bfx}{L},\;\bfy^*=\frac{\bfy}{L},\;\bfy_r^*=\frac{\bfy_r}{L},\;\bfy_{cm}^*=\frac{\bfy_{cm}}{L},\nonumber\\
&&M_r^*=\frac{M_r}{\rho_r  L^3},\;\calI_r^*=\frac{\calI_r}{\rho_rL^5},\;\bfg^*=\frac{\bfg
}{g},\;\bfw^*=\frac{\bfw}{2\pi f},\;\bfT^*=\frac{\bfT}{E}=\bar\bfT,\nonumber\\
&&\bff_1^*=\frac{\bff_1}{\rho_f L^2V^2},\;\bff_2^*=\frac{\bff_2}{EL^2},\;\bff_3^*=\frac{\bff_3}{\rho_rg L^3}=M_r^*\bfg^*,\;\bar\bfm_1^*=\frac{\bar\bfm_1}{\rho_f L^3V^2},\;\bar\bfm_2^*=\frac{\bar\bfm_2}{EL^3},\label{dimensionless_variables}
\eeqs
where the superscript $*$ represents the dimensionless quantities, and the definitions of $\bff_1,\bff_2$ and $\bff_3$ are given below in (\ref{fs}).  Then the dimensionless forms of equations of motion are found to be
\beq
\renewcommand{\arraystretch}{1.5}
\left\{\begin{array}{rcll}
\bfv^*_{t^*} + (\nabla^*\bfv^*)\, \bfv^* &=& -\nabla^* P^* + \frac{1}{Re}\Delta^* \bfv^*+\frac{1}{Fr^2} \bfg^*,\quad {\rm div}\ \bfv^* = 0,&{\rm for \ fluids},\\
\ddot{\bfy}^* &=&\frac{1}{Em}{\rm div} _{\bfx^*}\bfT^*+\frac{1}{Fr^2}\bfg^*,&{\rm for \ elastic \ solids},\\
M_r^*\ddot{\bfy}_{cm}^*  &=& \frac{\rho_f}{\rho_r}\bff_1^*+\frac{\rho}{\rho_r}\frac{1}{Em}\bff_2^*+\frac{1}{Fr^2}\bff_3^*,\quad {\rm and}\\
\calI_r^*\dot{\bfw}^*  &=& - \mathsmaller{2\pi St}\ \bfw^* \times\calI_r^* \bfw^* +\frac{1}{2\pi St}\bfR^T(\frac{\rho_f}{\rho_r}\bar\bfm_1^*+\frac{\rho}{\rho_r}\frac{1}{Em}\bar\bfm_2^*),&{\rm for \ rigid \ bodies},
\end{array}\right.\label{dimensionless_equations}
\eeq
where
\beq
Re=\frac{VL}{\nu},\quad Fr=\frac{V}{\sqrt{g L}},\quad Em=\frac{\rho V^2}{E},\quad St=\frac{fL}{V}.
\label{dless}
\eeq
For an organism in a fluid environment modeled as a system of  elastic solids, rigid bodies, and fluids, if the dimensionless numbers $Re, Fr, Em$, $St$ and the two density ratios $\frac{\rho_f}{\rho_r}$ and $\frac{\rho}{\rho_r}$ are the same, then the satisfaction of the equations of motion of Organism 1 implies that they are satisfied for Organism 2.  Matching the two density ratios gives immediately the scalings for density in Table \ref{table1}. Also, the scalings for geometry, time and density, together with the equality of these dimensionless numbers, imply the scalings listed in Table \ref{table1} for Young's modulus, the shear modulus, the kinematic viscosity and the shear modulus.

We note that there are equivalent sets of scaling laws that involve matching
other dimensionless numbers besides $Re$, $Fr$, $Em$ and $St$. For example,
matching $Fr$ and $Em$ for Organisms 1 and 2 implies that the dimensionless number $E/\rho f^2 L^2$
is the same for the two organisms.  Although matching of $E/\rho f^2 L^2$  is 
arguably more relevant for periodic motions (e.g., flying, swimming, running) it
adds no new information to Table \ref{table1}.

\vspace{5mm}
\noindent{\large \bf Boundary and interaction conditions  }
\vspace{2mm}

\noindent If certain boundary conditions are satisfied for Organism 1, corresponding conditions must necessarily be satisfied for Organism 2.
We choose accepted forms of these conditions, namely 
a) the elastic body satisfies traction conditions where it meets the fluid and displacement conditions where it meets the rigid body; b) the overall force and moment on the rigid body arises from the traction produced by the fluid and elastic bodies,
and the gravitational body force;  c) the fluid satisfies a (generally time dependent) no-slip condition where it meets the elastic and rigid bodies.

In the items below we assume that we have matched $Re$, $Fr$, $Em$, $St$ 
and the two density ratios as described in the preceding subsection. 
The gravitational body force is 
included (above) as a term in the
equations of motion and therefore has already been included. The reference configuration for the rigid body is $\Omega_r$, for the stiff elastic body is $\Omega$, for the soft elastic body is $\Omega_s$, and the interfaces between solids are denoted by $\calS_1=\partial\Omega_r\cap\partial\Omega, \calS_2=\partial\Omega_r\cap\partial\Omega_s, \rm{and}\, \calS_3=\partial\Omega_s\cap\partial\Omega$.

\begin{enumerate}[leftmargin=*]
\item {\bf Boundary conditions on the elastic bodies} due to the fluid and rigid body.

The Cauchy stress in the fluid at the boundaries $\partial\Omega\setminus(\calS_1 \cup \calS_3)$ (stiff elastic material) or $\partial\Omega_s\setminus(\calS_2 \cup \calS_3)$ (soft elastic material) is
\beq
\left.\begin{array}{c}
\boldsymbol \sigma_{so}(\bfx,t)\bfn  \\
{\rm or}\\
\boldsymbol \sigma_{st}(\bfx,t)\bfn
\end{array}\right\}
 =\boldsymbol \sigma_f\bfn= \rho_f \bigg( - p(\bfy(\bfx,t),t) \bfI +  \nu \big(\nabla_{\bfy} \bfv(\bfy(\bfx,t),t) + (\nabla_{\bfy}  \bfv(\bfy(\bfx,t),t)^T \big) \bigg) \bfn,
\label{tract1s}
\eeq
where  $\bfn$ is the unit outward surface normal of $\bfy(\calS_1)$ and $\bfy_s(\calS_2)$ for the corresponding cases. It follows from this formula, geometric similarity (so the corresponding outward normals are the same), the scaling of
density, and the equality of $Re$  and $Em$ numbers that satisfaction of traction conditions on the elastic body from the fluid for Organism 1 implies its satisfaction for Organism 2.  Assuming displacement conditions hold
for Organism 1 at the boundary between the elastic and rigid bodies, then they hold at this boundary for Organism 2 by geometric similarity, i.e., by the scaling of $\bfx, \bfy$ in (\ref{dimensionless_variables}).

The rigid body imposes displacement conditions on the elastic bodies:
\beq
\left.\begin{array}{c}
\bfy_{st}(\bfx,t)  \\
{\rm and}\\
\bfy_{so}(\bfx,t)
\end{array}\right\}
 =\bfy_r(\bfx,t)=\bfR(t)(\bfx-\bfx_{cm})+\bfy_{cm}(t),
\eeq
where $\bfx\in\calS_1$ for stiff elastic solid and $\bfx\in\calS_2$ for soft elastic solid. Again, by geometric similarity, if these conditions hold for Organism 1 at time $t$ then they hold for Organism 2 at time $\alpha t$. For the same reason this is also true for boundary conditions for the stiff elastic body interacting with the soft elastic body, i.e.,
\beq
\boldsymbol \sigma_{st}(\bfx,t)\bfn=\boldsymbol \sigma_{so}(\bfx,t)\bfn,\quad\bfy_{st}(\bfx,t)=\bfy_{so}(\bfx,t)\quad\bfx\in\calS_3.
\eeq
Here $\bfn$ represents the outward normal to $\bfy_{st}(\calS_3)$.

    \item {\bf Interaction conditions for the rigid body} due to the fluid and elastic solids. \\
The Cauchy stress in the fluid at the boundary ($\partial\Omega_r\setminus\calS$) with the rigid body, expressed in Lagrangian variables $(\bfx,t)$, is
\beq
\boldsymbol \sigma_f(\bfx,t) = \rho_f \bigg( - p(\bfy_r(\bfx,t),t) \bfI +  \nu \big(\nabla_{\bfy} \bfv(\bfy_r(\bfx,t),t) + (\nabla_{\bfy}  \bfv(\bfy_r(\bfx,t),t)^T \big) \bigg).\label{cauchy}
\eeq

Then, the force and moment on the rigid body from the fluid 
and elastic solids are,
\beqs
\bff_r(t) &=& \int_{\partial \Omega_r \setminus \calS} \boldsymbol \sigma_f(\bfx,t) \bfR \bfn_0\, da_0 + \int_{\calS} \bfT \bfn_0\, da_0+M_r\bfg, \nonumber \\
&=:&\bff_1+\bff_2+\bff_3\label{fs}\\
\bfm_r(t) &=& \int_{\partial \Omega_r \setminus \calS}\bfy_r(\bfx,t) \times \boldsymbol \sigma_f(\bfx,t) \bfR \bfn_0\, da_0 + \int_{\calS} \bfy_r \times \bfT \bfn_0\, da_0+\bfy_{cm}\times M_r\bfg.\label{ms}
\eeqs
where $\bff_1,\ \bff_2,$ and $ \bff_3$ represent the force on the rigid body from the fluid, elastic solid, and gravitational body force, respectively; $\calS(=\calS_1+\calS_2)$ represents the boundary between the rigid body and the elastic bodies in the reference domain: $\bfn_0$ is the unit outward surface normal of $\calS$.
By the scaling of density and the matching of $Em$, 
$Fr$ and $St$ (see (\ref{dimensionless_equations})),
if (\ref{fs}) and (\ref{ms}) give the force and moment on the rigid body at time $t$ for Organism 1, then they
give the force and moment at $\alpha t$ for Organism 2.

\item {\bf Boundary conditions on the fluid} due to rigid body, elastic solid, and the gravitational body force.\\
The fluid satisfies the no-slip conditions:
\beqs
\bfv(\bfy_{st}(\bfx, t),t) &=& \dot{\bfy}_{st}(\bfx,t), \ \  \bfx \in \partial \Omega \setminus (\calS_1\cup\calS_3),\\
\bfv(\bfy_{so}(\bfx, t),t) &=& \dot{\bfy}_{so}(\bfx,t), \ \  \bfx \in \partial \Omega_s \setminus (\calS_2\cap\calS_3),  \quad {\rm and} \\
\bfv(\bfy_r(\bfx, t),t) &=& \dot{\bfy}_r(\bfx,t),\ \ \bfx \in \partial \Omega_r \setminus (\calS_1\cup\calS_2).
\eeqs
It directly follows from the scaling laws for geometry,
time, motion and velocity (Table \ref{table1}) that satisfaction of these condition for Organism 1 implies their satisfaction for Organism 2. 
At infinity $|\bfr|\to\infty$, the fluid satisfies
\beq
\bfv(\bfr,t)\to\bfv_0\ {\rm and}\ \boldsymbol \sigma_f(\bfr,t)=-\rho_f p(\bfr,t)\bfI,
\eeq
for both Organisms, where $\bfv_0$ is the appropriate free stream velocity at infinity.
\end{enumerate}

\vspace{5mm}
\noindent{\large \bf Example of the use of Table \ref{table1}}
\vspace{2mm}

\noindent 
As an example of the use of Table \ref{table1}, consider the first result 
(\ref{scale_fish}) of the paper relating frequency $f$, length $L$ and mass $M$
of geometrically similar small (superscript $(s)$) and 
large (superscript $(\ell)$) organisms:
\beq
\frac{f^{(\ell)}}{f^{(s)}} =  \left[\frac{L^{(\ell)}}{L^{(s)}}\right]^{-\frac{1}{2}}\!= \left[\frac{M^{(\ell)}}{M^{(s)}} \right]^{-\frac{1}{6}}. \label{3ratios}
\eeq
Without loss of generality we assume
Organism 1 is the small, and 
Organism 2 is the large organism.
Referring to Table \ref{table1}, all cases considered are on the earth or at modest
elevation, so the gravity force is 
considered to be the same.  Hence, from
the ``Gravity'' entry of Table \ref{table1}, we have
\beq
\lambda = \alpha^2, \label{lamal}
\eeq
and the density is the same, so $\eta = 1$.  Hence, from the entries for ``Geometry'',``Mass'' and ``Frequency'',
and (\ref{lamal}), we have 
\beq
\frac{L^{(\ell)}}{L^{(s)}} = \lambda, \quad \frac{M^{(\ell)}}{M^{(s)}} = \lambda^3, \quad \frac{f^{(\ell})}{f^{(s)}} = \frac{1}{\alpha} = \sqrt{\frac{1}{\lambda}},
\eeq
which gives (\ref{3ratios}).  Scaling laws for other quantities can
be deduced from Table \ref{table1}, but there are
no further restrictions on the ratios
given in (\ref{3ratios}).  Of course, if
we alternatively identify Organism 1 as
the large organism and Organism 2 as the small
organism, the same results are obtained.

\vspace{7mm}
\noindent{\LARGE \bf Database}
\vspace{2mm}

\noindent In this extensive curated database we report all the raw data that is used in the main text,
and we provide metadata appropriate for 
discussions of scaling laws.  All fitting of 
graphs is done by direct regression of the measured data, and does not use any information about 
scaling laws.

\vspace{3mm}
\noindent{\bf Birds}

\begin{figure}[H]
\centering
\includegraphics[width=\textwidth]{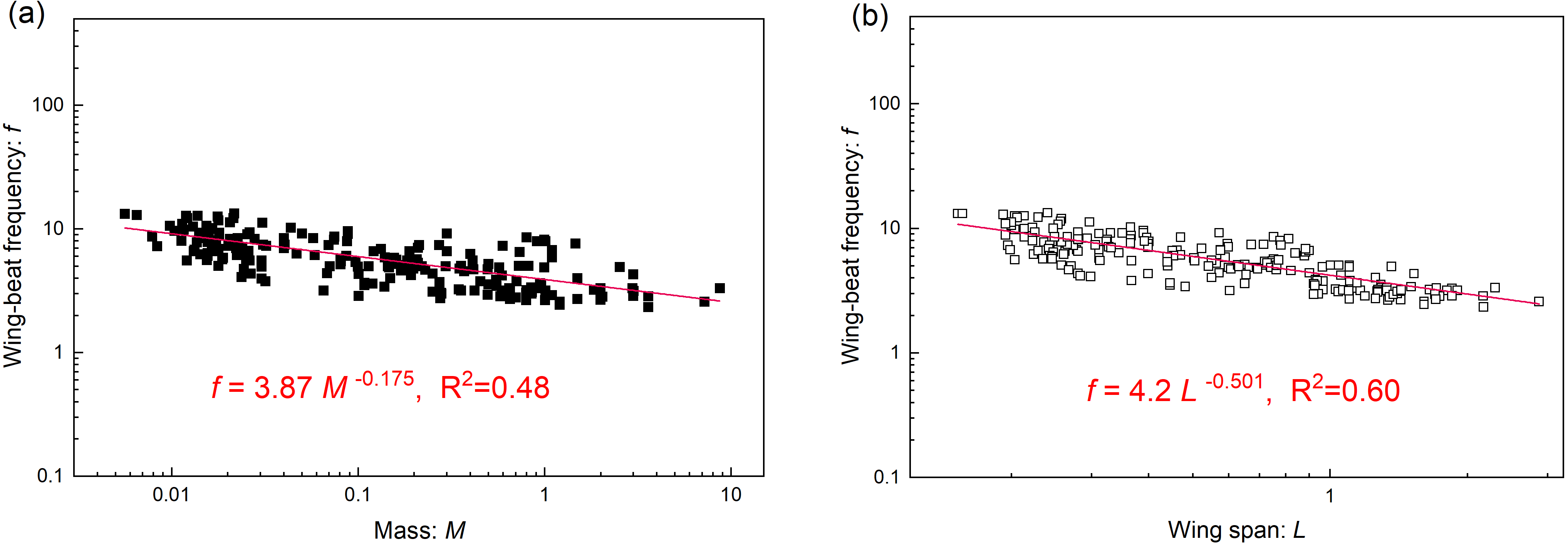}
\caption{Data reported by Bruderer et al.\cite{bruderer2010wing} for 200 bird species.}
\label{s1}
\end{figure}

\vspace{2mm}
In Figure 1(A,B), we collect the data of birds from Bruderer et al. \cite{bruderer2010wing}, measured by using radar, camera recordings, and videos, for 153 western Palaearctic, 2 African, and  45 Palaearctic species. Specifically,
\begin{itemize}[leftmargin=*]
    \item the observed birds included {\it Waders, Waterfowl, Herons, Soaring birds, Gulis $\&$ terns, Pigeons $\&$ sandgrouse, Near-passerines, Corvids, Falcons, Swifts, Swallows $\&$ martins, and  Passerine-type birds};
    \item the species measured by the authors included {\it Great Crested Grebe Podiceps cristatus,  Cormorant Phalacrocorax carbo,  White Pelican Pelecanus onocrotalus,  Night Heron Nycticorax nycticorax,  Little Egret Egretta garzetta,  Grey Heron Ardea cinerea, Purple Heron Ardea purpurea, Black Stork Ciconia nigra, White Stork Ciconia ciconia, Spoonbill Platalea leucorodia, Flamingo Phoenicopterus ruber, Mute Swan Cygnus olor, Gadwall Anas strepera, Mallard Anas platyrhynchos, Pintail Anas acuta, Shoveler Anas clypeata, Pochard Aythya farina, Tufted Duck Aythya fuligula, Goosander Mergus merganser, Honey Buzzard Pernis apivorus, Black Kite Milvus migrans, Red Kite Milvus milvus, African Fish Eagle Haliaeetus vocifer, Egyptian Vulture Neophron percnopterus, Marsh Harrier Circus aeruginosus, Hen Harrier Circus cyaneus, Pallid Harrier Circus macrourus, Montagu's Harrier Circus pygargus, Goshawk Accipiter gentilis, Sparrowhawk Accipiter nisus, Levant Sparrowhawk Accipiter brevipes, Gabar Goshawk Micronisus gabar, Common Buzzard Buteo buteo buteo, Steppe Buzzard Buteo buteo vulpinus, Lesser Spotted Eagle Aquila pomarine, Booted Eagle Hieraaetus pennatus, Bonelli's Eagle Hieraaetus fasciatus, Osprey Pandion haliaetus, Lesser Kestrel Falco naumanni, Kestrel Falco tinnunculus, Red-footed Falcon Falco vespertinus, Hobby Falco Subbuteo, Eleonora's Falcon Falco eleonorae, Sooty Falcon Falco concolor, Lanner Falco biarmicus, Peregrine Falcon Falco peregrinus, Barbary Falcon Falco pelegrinoides, Quail Coturnix coturnix, Coot Fulica atra, Oystercatcher Haematopus ostralegus, Stilt Himantopus himantopus, Cream-coloured Courser Cursorius cursor, Collared Pratincole Glareola pratincola, Black-winged Pratincole Glareola nordmanni, Grey Plover Pluvialis squatarola, Lapwing Vanellus vanellus, Sanderling Calidris alba, Dunlin Calidris alpina, Ruff Philomachus pugnax, Snipe Gallinago gallinago, Bar-tailed Godwit Limosa lapponica, Curlew Numenius arquata, Greenshank Tringa nebularia, Green Sandpiper Tringa ochropus, Little Gull Larus minutus, Black-headed Gull Larus ridibundus, Common Gull Larus canus, Lesser black-backed Gull Larus fuscus, Yellow-legged Gull Larus cachinnans, Black Tern Chlidonias niger, White-winged Black Tern Chlidonias leucopterus, Crowned Sandgrouse Pterocles coronatus, Black-bellied Sandgrouse Pterocles orientalis, Rock Dove Columba livia, Wood Pigeon Columba palumbus, Turtle Dove Streptopelia turtur, Long-eared Owl (Asio otus), Short-eared Owl Asio flammeus, Red-necked Nightjar Caprimulgus ruficollis, Alpine Swift Apus/Tachymarptis melba, Swift Apus apus, Pallid Swift Apus pallidus, Little Swift Apus affinis, Bee-eater Merops apiaster, Hoopoe Upupa epops, Wryneck Jynx torquilla, Crested Lark Galerida cristata, Woodlark Lullula arborea, Skylark Alauda arvensis,  Sand Martin Riparia riparia, Crag Martin Ptyonoprogne/Hirundo rupestris,  Barn Swallow Hirundo rustica,  Red-rumped Swallow Hirundo daurica, House Martin Delichon urbica, Tree Pipit Anthus trivialis, Meadow Pipit Anthus pratensis,  Water Pipit Anthus spinoletta,   Yellow Wagtail Motacilla flava, Grey Wagtail Motacilla cinerea, Pied Wagtail Motacilla alba, Dunnock Prunella modularis, Robin Erithacus rubecula, Nightingale Luscinia megarhynchos, Black Redstart Phoenicurus ochruros,   Redstart Phoenicurus phoenicurus, Whinchat Saxicola rubetra,  Wheatear Oenanthe oenanthe, Ring Ouzel Turdus torquatus,   Blackbird Turdus merula,   Fieldfare Turdus pilaris,  Song Thrush Turdus philomelos,  Mistle Thrush Turdus viscivorus, Marsh Warbler Acrocephalus palustris, Reed Warbler Acrocephalus scirpaceus,  Great Reed Warbler Acrocephalus arundinaceus, Grasshopper Warbler Locustella naevia, Icterine Warbler Hippolais icterina, Melodious Warbler Hippolais polyglottal,  Olivaceous Warbler Hippolais pallida, Orphean Warbler Sylvia hortensis,   Lesser Whitethroat Sylvia curruca, Whitethroat Sylvia communis, Garden Warbler Sylvia borin, Blackcap Sylvia atricapilla, Willow Warbler Phylloscopus trochilus,  Goldcrest Regulus regulus,   Firecrest Regulus ignicapillus, Spotted Flycatcher Muscicapa striata, Pied Flycatcher Ficedula hypoleuca,  Coal Tit Parus ater,  Red-backed Shrike Lanius collurio, Jay Garrulus glandarius,  Magpie Pica pica, Nutcracker Nucifraga caryocatactes, Jackdaw Corvus monedula, Rook Corvus frugilegus, Carrion Crow Corvus corone, Brown-necked Raven Corvus ruficollis, Raven Corvus corax, Starling Sturnus vulgaris, Chaffinch Fringilla coelebs, Serin Serinus serinus,  Greenfinch Carduelis chloris, Goldfinch Carduelis carduelis, Siskin Carduelis spinus, Linnet Carduelis cannabina, {\rm and} Ortolan Bunting Emberiza hortulana}; and
    \item the species collected from other literature included {\it Black-browed Albatross Diomedea melanophris, Wandering Albatross Diomedea exulans, Giant Petrel Macronectes giganteus/halli, Fulmar Fulmarus glacialis, Wilson’s Storm-petrel Oceanites oceanicus, Gannet Morus bassanus, Double-crested Cormorant Phalacrocorax auritus, Shag Phalacrocorax aristotelis, Magnificent Frigatebird Fregata magnificens, Bewick’s Swan Cygnus columbianus, Wooper Swan Cygnus cygnus, Pink-footed Goose Anser brachyrhynchus, Greylag Goose Anser anser, White-fronted Goose Anser albifrons, Barnacle Goose Branta leucopsis, Brent Goose Branta bernicla, Shelduck Tadorna tadorna, Wigeon Anas penelope, Eider Somateria mollissima, Wood Sandpiper Tringa glareola, Common Sandpiper Actitis hypoleucos,  Turnstone Arenaria interpres, Arctic Skua Stercorarius parasiticus, Great Skua Cataracta skua, Herring Gull Larus argentatus, Great Black-backed Gull Larus marinus, Kittiwake Rissa tridactyla, Gull-billed Tern Gelochelidon nilotica, Royal Tern Sterna maxima,   Sandwich Tern Sterna sandvicensis,   Common Tern Sterna hirundo, Guillemot Uria aalge, Razorbill Alca torda, Puffin Fratercula arctica, Cuckoo Cuculus canorus, Thrush Nightingale Luscinia luscinia, Redwing Turdus iliacus, Sege Warbler Acrocephalus schoenobaenus, Chiffchaff Phylloscopus collybita, Brambling Fringilla montifringilla, {\rm and} Redpoll Carduelis flammea.} 
 \end{itemize}
 See Figure \ref{s1} and see reference\cite{bruderer2010wing} for species details.
 
\begin{figure}[H]
\centering
\includegraphics[width=\textwidth]{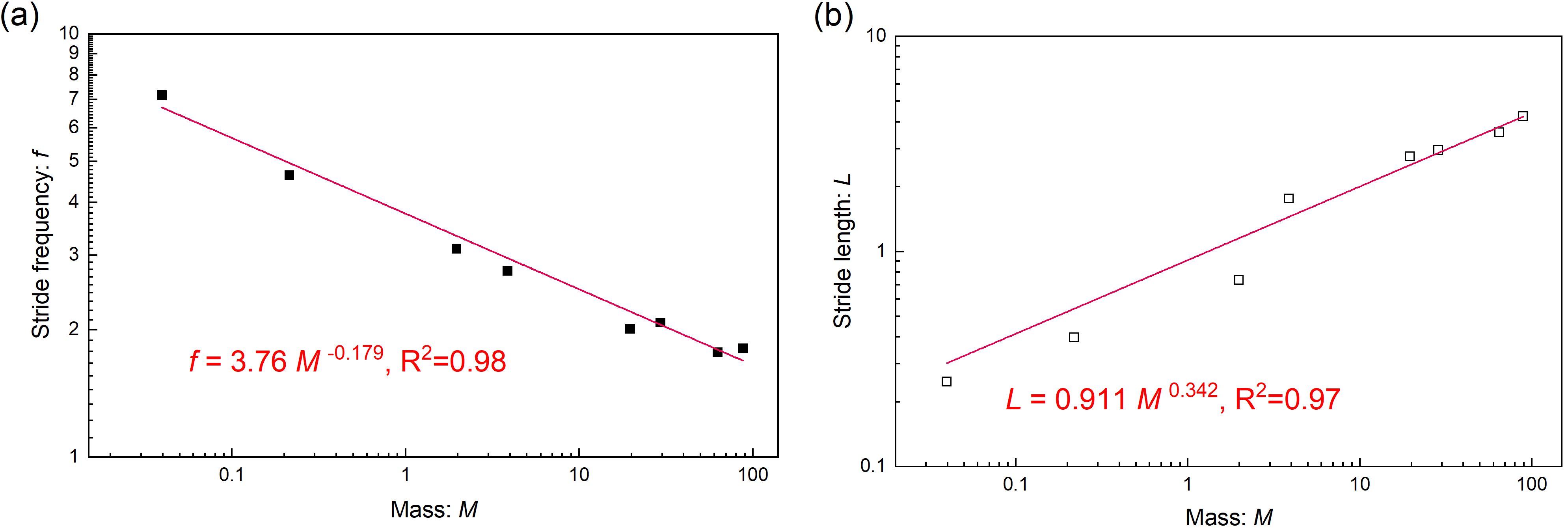}
\caption{Data reported by Gatesy and Biewener\cite{gatesy1991bipedal} for bipedal animals including human, painted quail, bobwhite, guineafowl, turkey, emu, rhea, and ostrich.}
\label{s2}
\end{figure}

\vspace{2mm}
We collect data reported by Gatesy and Biewener \cite{gatesy1991bipedal} on the locomotion of bipedal animals including {\it Painted quail, Bobwhite, Guineafowl, Turkey, Emu, Rhea, and Ostrich}. We summarize the data on stride frequency vs.~body mass in Figure 1(A). The stride length was found to scale as $L\propto M^{0.342}$ with $M$ the body mass, which is close to our prediction ($L\propto M^{1/3}$) from (\ref{scale_fish}) within the acceptable error. The data was measured at the animal's fastest treadmill speed. The authors also pointed out that for the larger species the speed they recorded was unlikely to be the fastest-running
speeds, which might affect the scaling parameters. See Figure \ref{s2}.

\begin{figure}[H]
\centering
\includegraphics[width=\textwidth]{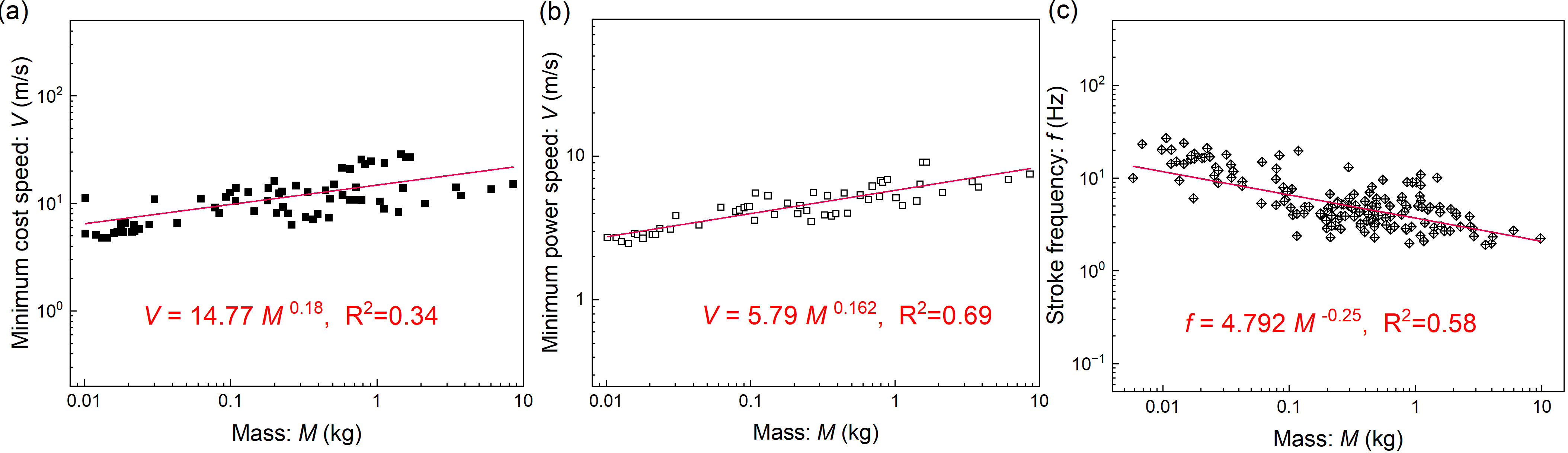}
\caption{Data reported by Rayner\cite{rayner1979new} for birds.}
\label{s3}
\end{figure}

\vspace{2mm}
We collected the data reported by Rayner \cite{rayner1979new} about stroke period and flight speed vs.~body mass of birds including {\it perching birds, shorebirds, and ducks} in which the stroke period is converted to frequency by taking the reciprocal.  See Figure \ref{s3}.

\begin{figure}
\centering
\includegraphics[width=0.9\textwidth]{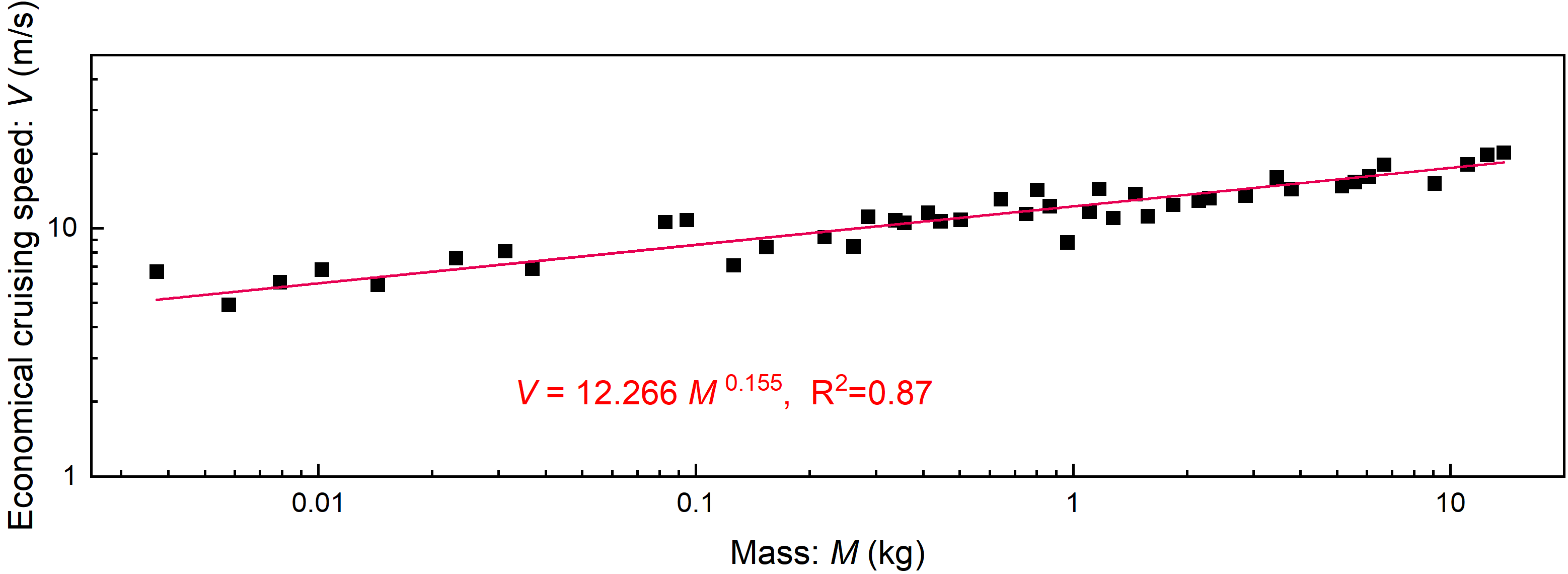}
\caption{Data reported by Greenewalt \cite{greenewalt1962dimensional} and \cite{pennycuick1969mechanics} for birds.}
\label{s4}
\end{figure}

We summarize the data measured by Greenewalt \cite{greenewalt1962dimensional} and collected by Pennycuick \cite{pennycuick1969mechanics} about the cruising speed vs.~body mass of flying animals including {\it Gymnogyps californianus, Diomedea exulans, Gypaetus barbatus, Ciconia ciconia, Phalacrocorax carbo, Bubo bubo, Fregata aquila, Larus argentatus, Milvus milvus, Falco peregrinus, Fulmarus glacialis, Strix aluco, Puffinus puffinus, Circus pygargus, Sterna hirundo, Apus apus, Cygnus cygnus, Ardeotis kori, Cygnus columbianus, Pseudogyps africanus, Leptoptilos crumeniferus, Anser anser, Sula bassana, Necrosyrtes manachus, Pteropus edulis, Anas platyrhynchos, Buteo buteo, Alca torda, Lagopus lagopus, Corvus corone, Columba livia, Scolopax rusticola, Fratercula arctica, Vanellus vanellus, Falco tinnunculus, Alle alle, Sturnus vulgaris, Passer domesticus, Parus major, Hirundo urbica, Troglodytes troglodytes, Phylloscopus bonellil, Vespertilio pipistrellus, Archilochus colubris}. The plot in \cite{pennycuick1969mechanics} shows the range of cruising speeds of different species. We extracted the middle point for each speed range.  See Figure \ref{s4}.

We summarize the data measured by Alerstam et al. \cite{alerstam2007flight} by using a tracking radar on bird fight for 138 species of six main Monophyletic groups including {\it swans, geese, ducks, flamingos, pigeons, swifts, waders, gulls, terns, divers, cormorants, pelican, herons, storks, crane, falcons, crows, songbirds, hawks, eagles, osprey, bee-eater}, etc.

\vspace{2mm}

\begin{figure}[H]
\centering
\includegraphics[width=\textwidth]{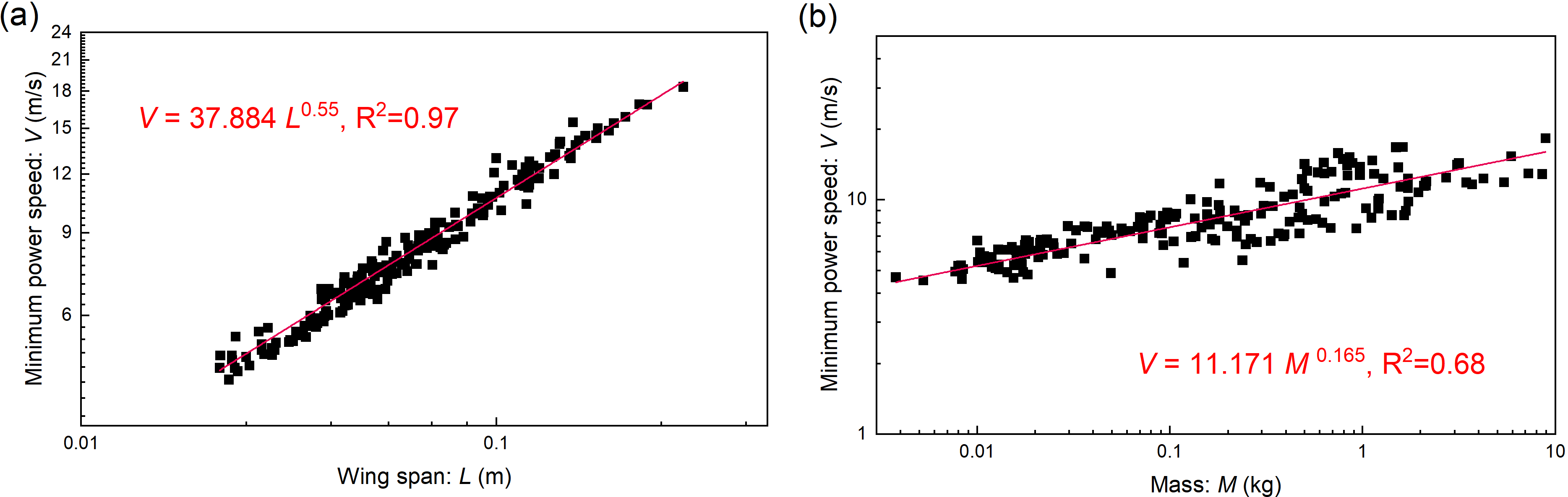}
\caption{Data reported by Greenewalt \cite{greenewalt1975flight} and Bejan et al.\cite{bejan2006unifying} for birds.}
\label{s5}
\end{figure}

The data reported by Greenewalt \cite{greenewalt1975flight} and collected by Bejan et al.~\cite{bejan2006unifying} for {\it perching birds, shorebirds,} and {\it ducks} concerns correlations between velocity, mass and span loading, and is extracted and summarized in our Figure 1(C) and (E). Span loading is defined as the total mass divided by the square of the wing span, which is proportional to length from (\ref{scale_fish}). We summarized the data about speed vs.~span loading in Figure 1(D), which is rephrased equivalently as speed vs.~length.  See Figure \ref{s5}.

\begin{figure}[H]
\centering
\includegraphics[width=\textwidth]{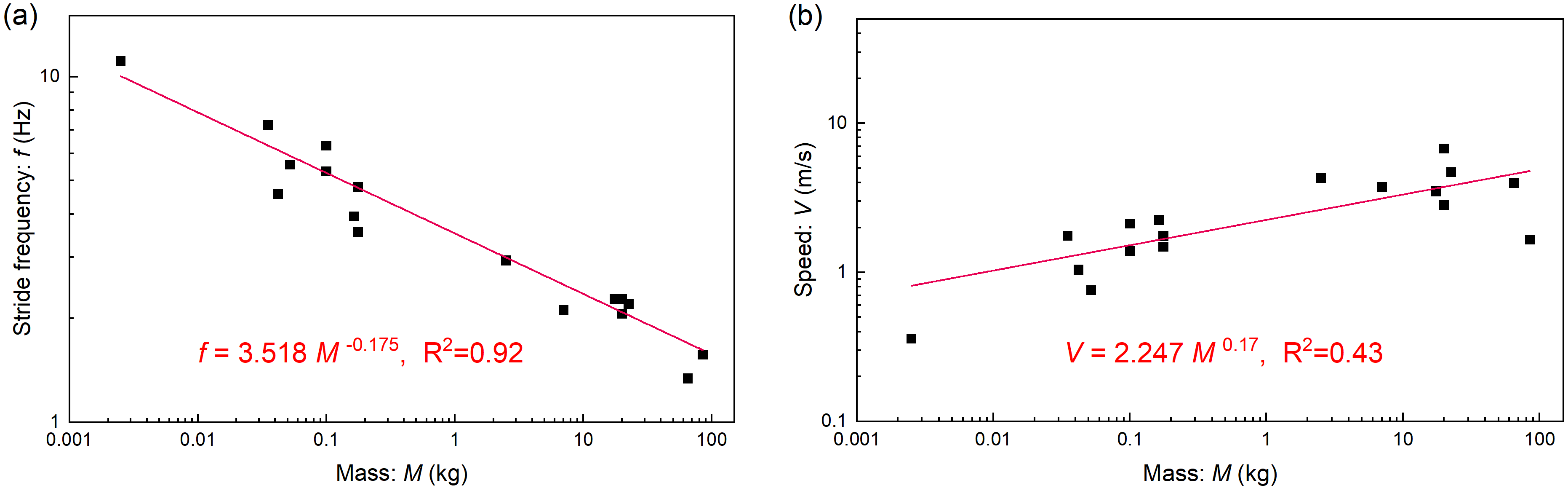}
\caption{Data reported by Blickhan and Full\cite{blickhan1993similarity} for birds.}
\label{s6}
\end{figure}

We collect the data collected by Blickhan and Full \cite{blickhan1993similarity} about the body mass, stride frequency, and speed in the bird species {\it Painted quail: Excalfactoria chinensis, Bobw. quail: Colinus virginianus, Turkey: Mellegria gallopavo, Rhea: Rhea americana, {\rm and} Springhare: Pedetes cafer}. See Figure \ref{s6}.
 
\begin{figure}[H]
\centering
\includegraphics[width=0.9\textwidth]{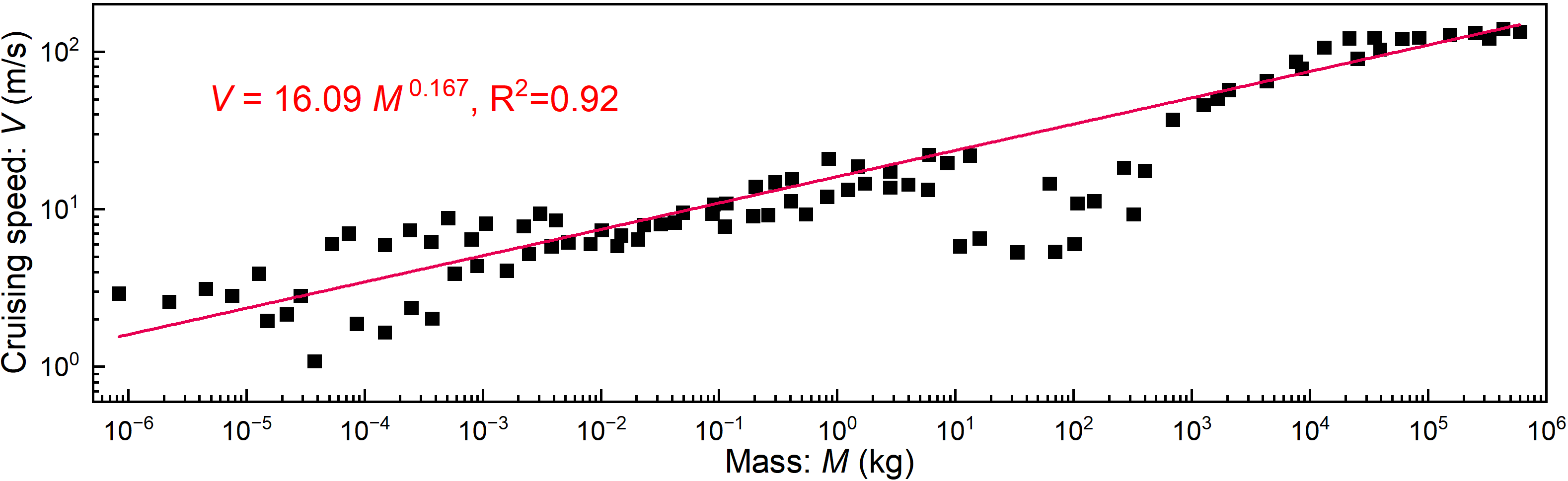}
\caption{Data collected by Tennekes\cite{tennekes2009simple} on birds' speed vs.~mass of animals. For completeness of the database, the data on insects from this reference are also included.}
\label{simple_science}
\end{figure}

We summarize the data collected by Tennekes \cite{tennekes2009simple} concerns the body mass and speed of birds including {\it Common tern, Dove prion, Black-headed gull, Black skimmer, Common gull, Kittiwake, Royal tern, Fulmar, Herring gull, Great skua, Great black-billed gull, Sooty albatross, Black-browed albatross, {\rm and} Wandering albatross}. See Figure \ref{simple_science}.

\vspace{2mm}
We include the data for penguins measured by Sato et al.~ \cite{sato2010scaling} about mean cruising speed and dominant stroke cycle frequency vs.~body mass. The species  include {\it Aptenodytes patagonicus, Pygoscelis papua, Pygoscelis adeliae, Pygoscelis antarcticus, Eudyptes chrysolophus, {\rm and}
Eudyptula minor}.

\vspace{3mm}
\noindent{\bf Fish} 

\begin{figure}[H]
\centering
\includegraphics[width=\textwidth]{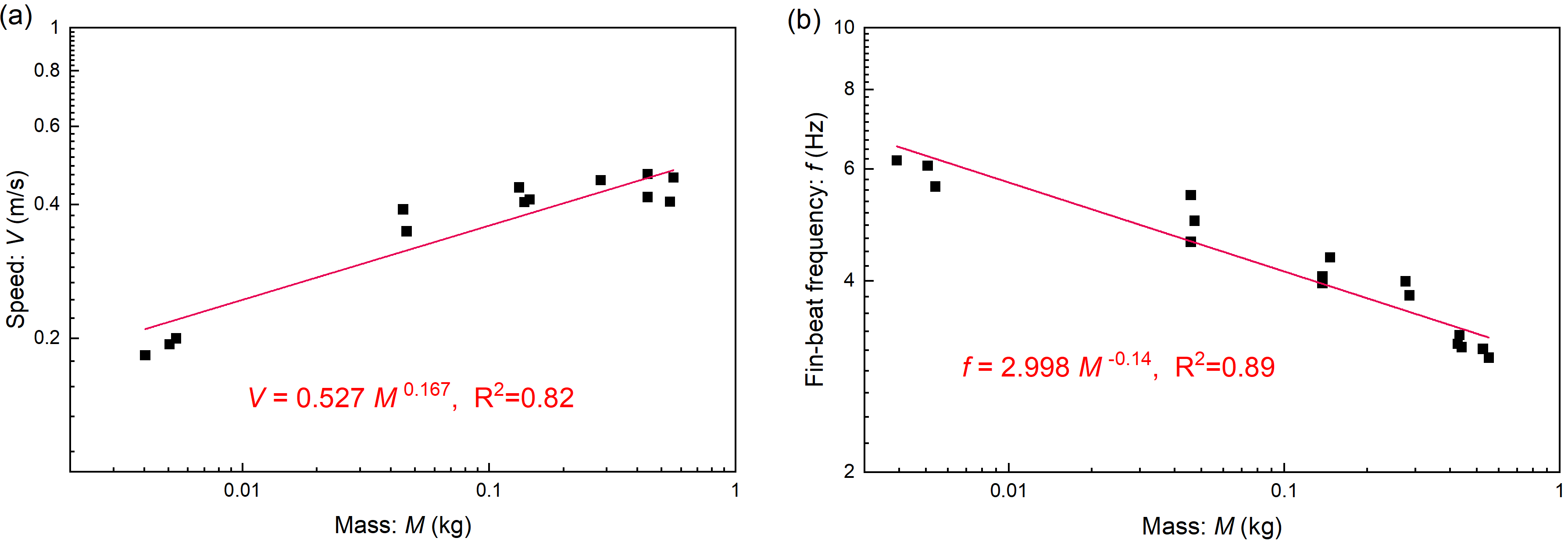}
\caption{Data reported by Drucker et al.\cite{drucker1996pectoral} for fish.}
\label{s8}
\end{figure}

\vspace{2mm}
We collect the data reported by Drucker et al.~\cite{drucker1996pectoral} on swimming fish including {\it Embiotoca lateralis, Cymatogaster aggregate, Pomoxis annularis, and Lepomis macrochirus}. The data was measured at the pectoral–caudal gait transition speed, and the fin-beat frequency vs.~body mass is summarized in Figure 1(A). See Figure \ref{s8}.

\begin{figure}[H]
\centering
\includegraphics[width=0.9\textwidth]{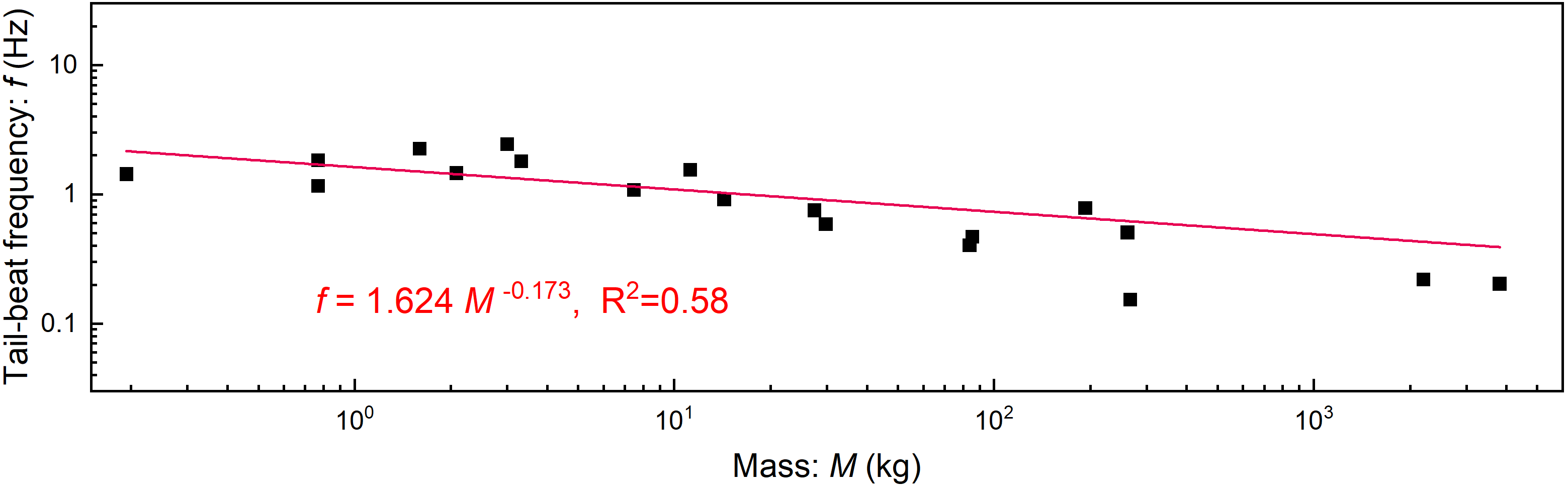}
\caption{Data collected by Watanabe \cite{watanabe2012slowest} for fish.}
\label{s9}
\end{figure}

We summarize the data collected by Watanabe \cite{watanabe2012slowest} from the literature on body mass, body length, swimming speed, and tail-beat frequency of fish, including {\it Atlantic cod, Gadus morhua, Atlantic herring, Clupea harengus, Basking shark, Cetorhinus Maximus, Blacktip reef shark, Carcharhinus melanopterus, Blue marlin, Makaira nigricans, Blue shark, Prionace glauca, Brown trout, Salmo trutta, Chinese sturgeon, Acipenser sinensis, Chinook salmon, Oncorhynchus tshawytscha, Chum salmon, Oncorhynchus keta, Greenland shark, Somniosus microcephalus, Japanese flounder, Paralichthys olivaceus, Lemon shark, Negaprion brevirostris, Nurse shark, Ginglymostoma cirratum, Ocean sunfish Mola mola, Pink salmon, Oncorhynchus gorbuscha, Scalloped hammerhead shark, Sphyrna lewini pups, Shortfin mako shark, Isurus oxyrinchus, Sockeye salmon, Oncorhynchus nerka, Tiger shark, Galeocerdo cuvier, Whale shark, Rhincodon typus, Whitetip reef shark, and Triaenodon obesus}. See Figure \ref{s9}.

Videler et al.~\cite{videler1991fish} measured the maximum swimming speed and the maximum tail-beat frequency of fish of different size, including the  species {\it Leuciscus leuciscus, Oncorhynchus mykiss, Carassius auratus, Trachurus symmetricus,  Scomber japonicus, Triakis henlei, Sardinops sagax, Acipenser fulvescens, Ginglymostoma cirratum, Triakis semifasciata, Negaprion brevirostris, Sphyrna tiburo, Carcharhinus melanoterus,  Carcharhinus leucas, Scomber scombrus, Pollachius virens, Gadus morhua, and Thunnus thynnus}.

\vspace{2mm}
Bainbridge \cite{bainbridge1958speed} collected extensive data from the literature and reported related data of his own.  This data includes speed of swimming fish as related to fish size, body mass, tail-beat frequency and tail-beat amplitude. We summarize the data on tail-beat frequency vs.~fish length in our Figure 1(B), for fish species of {\it Cyprinus carpio, Alburnus alburnus, Squalius cephalus, Salmo trutta, Scomber scombrus, Alosa finita, Perca perca, 
 Sciaena aquila, Gadus merlangus, G. luscus, Mugil capito, Scardinius erythrophthalmus, Trachinus vipera, Merluccius vulgaris, Esox lucius, Trigla pini,  Zeus faber, Micropterus salmoides, Trigla sp., Sebastes dactylopterus, Salmo fario, S. salar, Thunnus thynnus, Carassius auratus, Catostomus occidentalis, Carcharinus leucas, Protnicrops itaiara, Negaprion brevirostris, Sphyraena barracuda, Esox lucius, Scardinius erythroph-thaltnus, Leuciscus leuciscus, and Salmo irideus}.

\vspace{2mm}

The data collected by Rodr\'iguez et al.~\cite{sanchez2023scaling} concerns the tail beat frequency and swimming speed vs.~body length of fish, including the species  {\it Eptatretus stoutii, Myxine glutinosa, Petromyzon marinus, Taeniura lymma, Ginglymostoma cirratum, Triakis semifasciata, Negaprion brevirostris, Sphyrna tiburo, Carcharhinus limbatus, Carcharhinus leucas, Isurus oxyrinchus, Cetorhinus maximus, Carcharhinus melanopterus, Somniosus microcephalus, Negaprion brevirostris, Ginglymostoma cirratum, Sphyrna lewini, Isurus oxyrinchus, Galeocerdo cuvier, Rhincodon typus, Triaenodon obesus, Salmo irideus, Carassius auratus, Leuciscus leuciscus, Trachurus symmetricus, Clupea harengus, Salmo trutta, Acipenser sinensis, Oncorhynchus tshawytscha, Oncorhynchus keta, Paralichthys olivaceus, Mola mola, Oncorhynchus gorbuscha
Oncorhynchus nerka, Pollachius virens, Scomber scombrus, Istiophorus platypterus, Acipenser fulvescens, Salmo gairdneri, Sarda chiliensis
Acipenser fulvescens, Oncorhynchus nerka, Oncorhynchus tshawytscha, {\rm and} Protopterus annectens}.
\begin{figure}[H]
\centering
\includegraphics[width=0.9\textwidth]{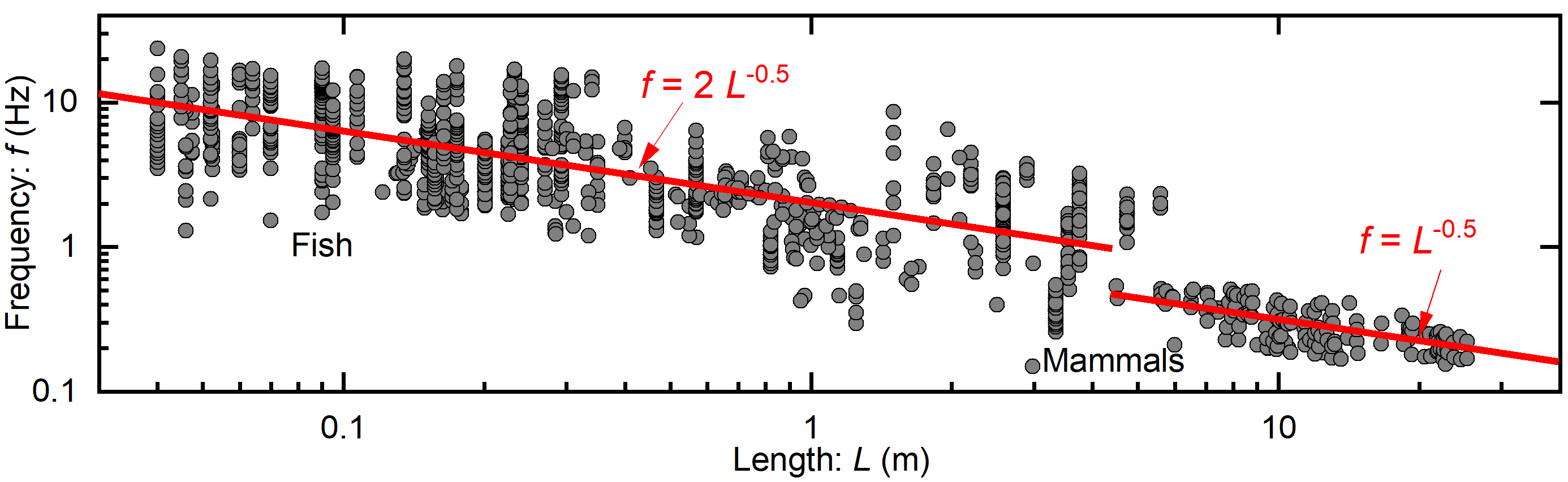}
\caption{Data reported Rodr\'iguez et al.~\cite{sanchez2023scaling} for swimming animals.}\label{frequency-length-fish}
\end{figure}
\vspace{3mm}

\noindent{\bf Reptiles}

\vspace{2mm}
The data reported by Garland \cite{garland1985ontogenetic} concerns the maximum running speed as a function of body mass for 60 agamid lizard {\it Amphibolurus nuchalis} including juveniles, adult males, adult females, gravid females, and long-term captive individuals.

\vspace{2mm}
The data reported by Marsh \cite{marsh1988ontogenesis} is about the stride length, stride frequency, and velocity vs.~body mass in the lizard {\it Dipsosaurus dorsalis} with a body temperature of $35^\circ$ and $40^\circ$.

\vspace{2mm}
The data collected by Rodr\'iguez et al.~\cite{sanchez2023scaling} concerns the tail beat frequency and swimming speed vs.~body length of reptiles including {\it Alligator mississippiensis, Nerodia f. pictiventris,
Elaphe g. guttata, {\rm and} Aptenodytes forsteri}.

\vspace{2mm}
The data for Colubrid snakes measured by Jayne \cite{jayne1985swimming} concerns the forward swimming speed and body mass. The species include {\it Constricting Colubrid snakes: Elaphe g. guttata {\rm and} Nonconstricting Colubrid snakes: Nerodia fasciata pictiventris}.

\vspace{3mm}
\noindent{\bf Amphibians}

\begin{figure}[H]
\centering
\includegraphics[width=0.9\textwidth]{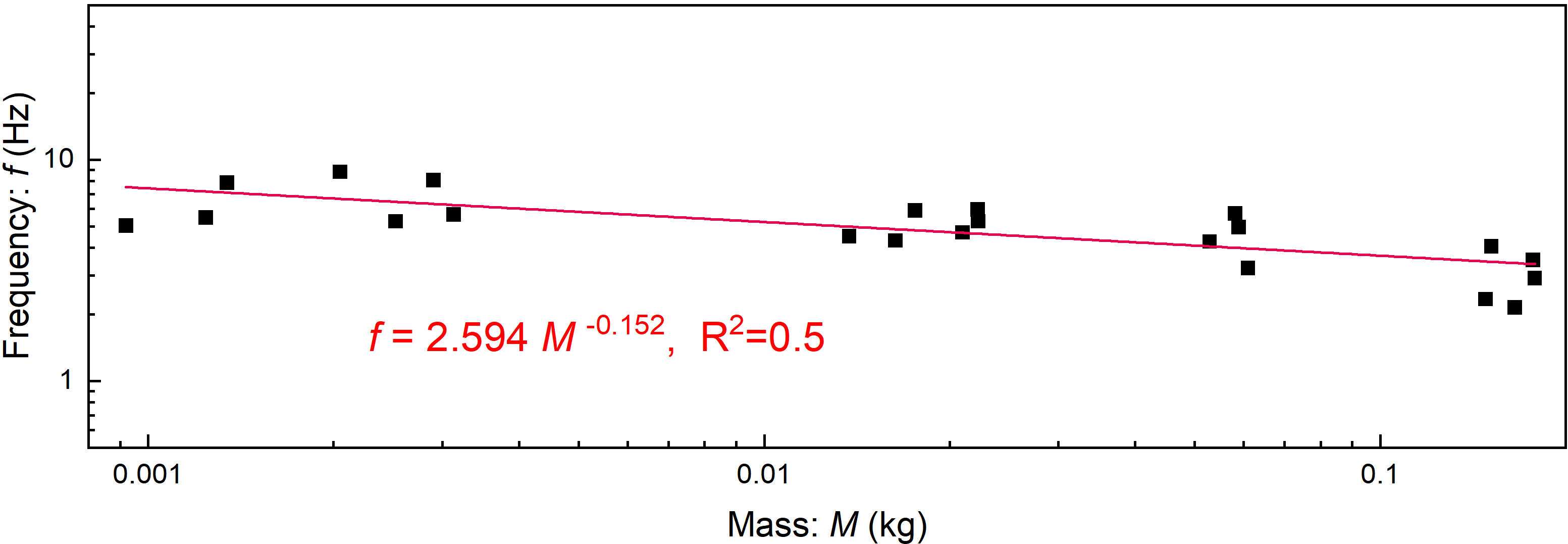}
\caption{Data reported by Clemente and Richards\cite{clemente2013muscle} for frogs.  }
\label{s10}
\end{figure}

\vspace{2mm}
Clemente and Richards \cite{clemente2013muscle} collected the maximum swimming speed and mean rotational velocity of the ankle as functions of body mass in 22 individual {\it Xenopus laevis}. The rotational velocity is converted to frequency by dividing $2\pi$. See Figure \ref{s10}.

The data collected by Rodr\'iguez et al.~\cite{sanchez2023scaling} includes the tail beat frequency and swimming speed vs.~body length of amphibians including {\it Rana septentrionalis, Rana clamitans, {\rm and} Rana catesbiana}.

\vspace{3mm}
\noindent{\bf Mammals}

\begin{figure}
\centering
\includegraphics[width=0.9\textwidth]{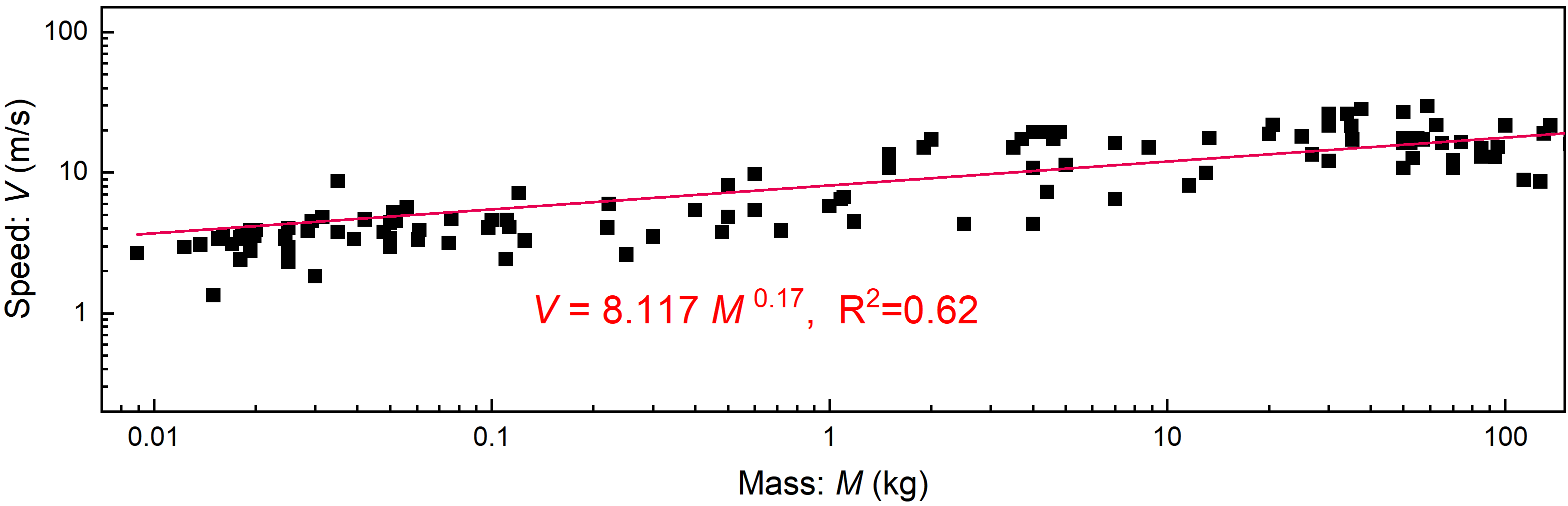}
\caption{Data reported by Iriarte-D{\'\i}az\cite{iriarte2002differential} for mammals.}
\label{s11}
\end{figure}

\vspace{2mm}
The data collected by Iriarte-D{\'\i}az \cite{iriarte2002differential} from various sources concerns body mass and maximum running speed of 144 terrestrial mammal species including {\it Ursus americanus, Acinonyx jubatus, Panthera pardus, Crocuta crocuta, Canis lupus, Hyaena hyaena, Canis familiarise, Lycaon pictus, Canis latrans, Meles meles, Canis aureus, Procyon lotor, Canis mesomelas {\rm or} adustus, Vulpes fulva, Nasua narica, Urocyon cinereoargenteus, Mephitis mephitis, Gorilla gorilla, Homo sapiens, Presbytis, Marmota monax, Uromys caudimaculatus, Sciurus niger, Spermophilopsis leptodactylus, Spermophilus undulatus, Spermophilus citellus, Sciurus carolinensis, Sciurus vulgaris {\rm and} persicus, Spermophilus beldingi, Rattus, Spermophilus saturates, Tamiasciurus hudsonicus, Spermophilus tridecemlineatus, Spermophilus tereticaudus, Neotoma lepida, Mesocricetus brandti, Tamias striatus, Dipodomys deserti, Ammospermophilus leucurus, Pseudomys nanus, Zymomys argurus, Dipodomys microps, Tamias amoenus, Microtus pennsylvanicus, Pseudomys australis, Heteromys dasmarestianus, Dipodomys ordii, Lyomis pictus, Chaetodipus baileyi,Dipodomys merriami,Notomys cervinus,Pitymys pinetorum,Tamias minimus, Zapus trinotatus, Peromyscus leucopus, Napeozapus insignis, Notomys alexis, Perognathus parvus, Peromyscus eremicus, Peromyscus truei, Onychomys torridus, Peromyscus maniculatus, Chaetodipus fallax, Zapus hudsonicus, Pseudomys hermannbergensis, Mus musculus, Leggadina forresti, Peromyscus crinitus, Microdipodops megacephalus, Perognathus longimembris, Lepus arcticus, Lepus alleni, Lepus europeus, Lepus townsendii, Lepus californicus, Oryctolagus cuniculus, Lepus americanus, Sylvilalagus, Macropus spp, Macropus eugenii, Bettongia penicilata, Potorus tridactylus, Isoodon obesulus, Dasyuroides byrnei, Monodelphis brevicaudata, Antechinus flavipes, Antechinus stuardii, Antechinomys laniger, Sminthopsis macroura, Sminthopsis crassicaudata, Cercatetus concinnus, Myrmecobius fasciatus, Loxodonta africana, Elephas maximus, Ceratotherium simum, Diceros bicornis, Equus caballus, Equus zebra, Tapirus bairdii, Equus burchelli, Equus hemionus, Hippopotamus amphibius, Giraffa camelopardalis, Bison bison, Bos sauveli, Syncerus caffer, Camelus dromedarius, Taurotragus oryx, Alces alces, Cervus elaphus, Connochaetes gnu, Hippotragus equinus, Connochaetes taurinus, Alcelaphus buselaphus, Damaliscus lunatus, Oreamnos americanus, Rangifer tarandus, Lama guanicoe, Ovis canadensis nelsoni, Phacochoerus aethiopicus, Odocoileus hemionus, Capra caucasia, Ovis ammon, Gazella granti, Odocoileus virginianus, Dama dama, Aepyceros melampus, Antilocapra americana, Capreolus capreolus, Rupicapra rupicapra, Antilope cervicapra, Saiga tatarica, Antidorcas marsupialis, Gazella subgutturosa, Procapra gutturosa, Capra aegagrus, Gazella thomsonii, Madoqua kirki, Ursus maritimus, Ursus maritimus, Panthera tigris, {\rm and} Panthera leo.} See Figure \ref{s11}.

\begin{figure}[H]
\centering
\includegraphics[width=\textwidth]{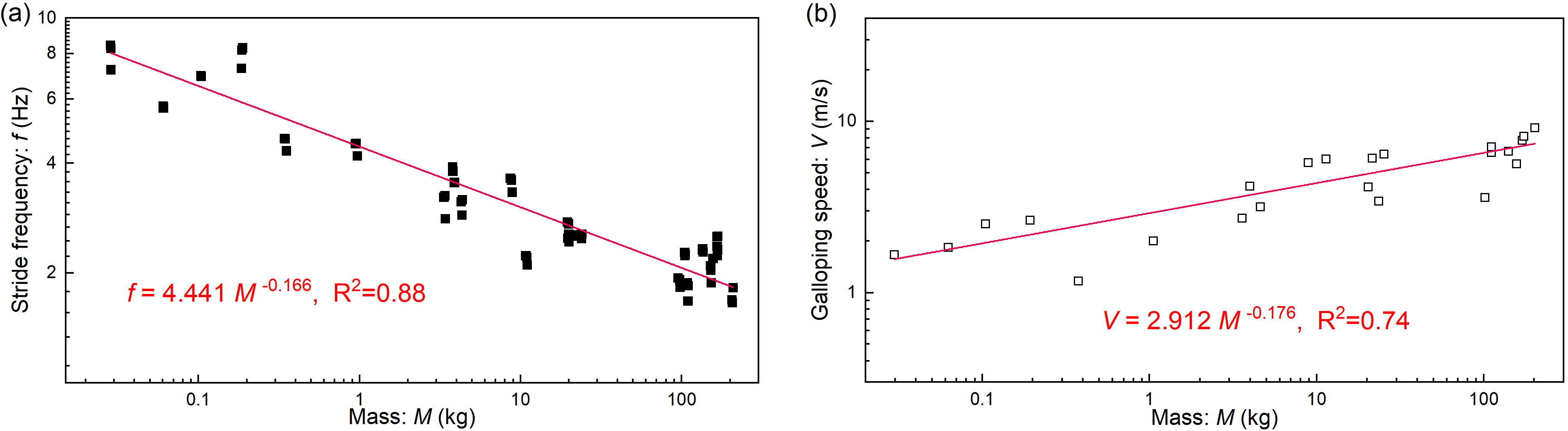}
\caption{Data reported by Heglund and Taylor 
 \cite{heglund1988speed,heglund1974scaling} for mammals.}\label{s12}
\end{figure}

The data reported by Heglund and Taylor \cite{heglund1988speed}  correlates stride frequency, speed, and body mass in mammals including {\it 13-lined ground squirrel, Suni, Dik-dik, Grant's gazelle, African domestic goat, Fat-tailed sheep, Wildebeest, Pony, Waterbuck, Pony, Zebu cattle, Donkey, Eland, White mouse, White rat, Dog, {\rm and} Horse} in which the data for last four kinds of mammals were measured in their previous work \cite{heglund1974scaling}. See Figure \ref{s12}.

In addition to the 144 species summarized by Iriarte-D{\'\i}az, Garland \cite{garland1983relation} collected the data on body mass and maximum running speed of 117 mammal species, and Garland and Janis \cite{garland1993does} compiled the data from a variety of sources of 49 mammal species. The species included in \cite{garland1983relation} and \cite{garland1993does} but not in \cite{garland1983relation} are {\it Ursus horribilis, Cervus canadensis, Tapirus americanus, Damaliscus korrigum, Lama guanacoe, Saiga  tatarica, Antidorcas  marsupialis, Thalarclos maritimus, Hyaena vulgaris, Erithizon dorsatum, Citellus undulatus, Citellus citellus, Citellus beldingi, Eutamius minimus, Talpa europaea, Scalopus aquaticus, Blarina brevicauda, Sylvilagus, Didelphis  marsupialis, Antechinomys spenceri, {\rm and} Bradypus tridactylus.}

\vspace{2mm}

\begin{figure}[H]
\centering
\includegraphics[width=\textwidth]{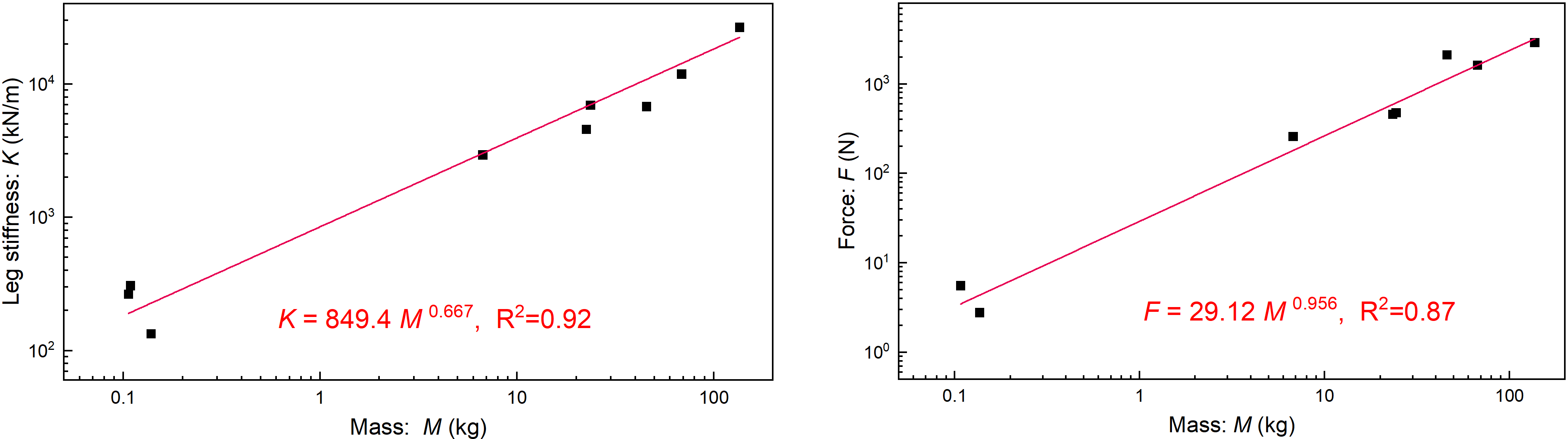}
\caption{Data reported by Farley et al.\cite{farley1993running} for legged animals.}\label{s13}
\end{figure}

The data reported by Farley et al. \cite{farley1993running} is about leg stiffness of {\it Dog: Canis familiaris, Horse: Equus caballus,  African pygmy goat: Capra hircus, Kangaroo: Megaleia rufa, Rat: Rattus norvegicus,  Rats: Dipodomys spectabilis, {\rm and} Tammar wallaby: Macropus eugenii}. Farley et al. modeled the legs of these animals as musculoskeletal spring systems and measured the stiffness of the ``leg springs" on a treadmill with a force platform equipped with strain gauges. The leg stiffness $K$ is formulated by $K=\frac{\Delta F}{\Delta L}$, where $\Delta F$ is the change in the force acting on the treadmill and $\Delta L$ is the change of leg height during the movement of the animals. See Figure \ref{s13}.

\begin{figure}[H]
\centering
\includegraphics[width=\textwidth]{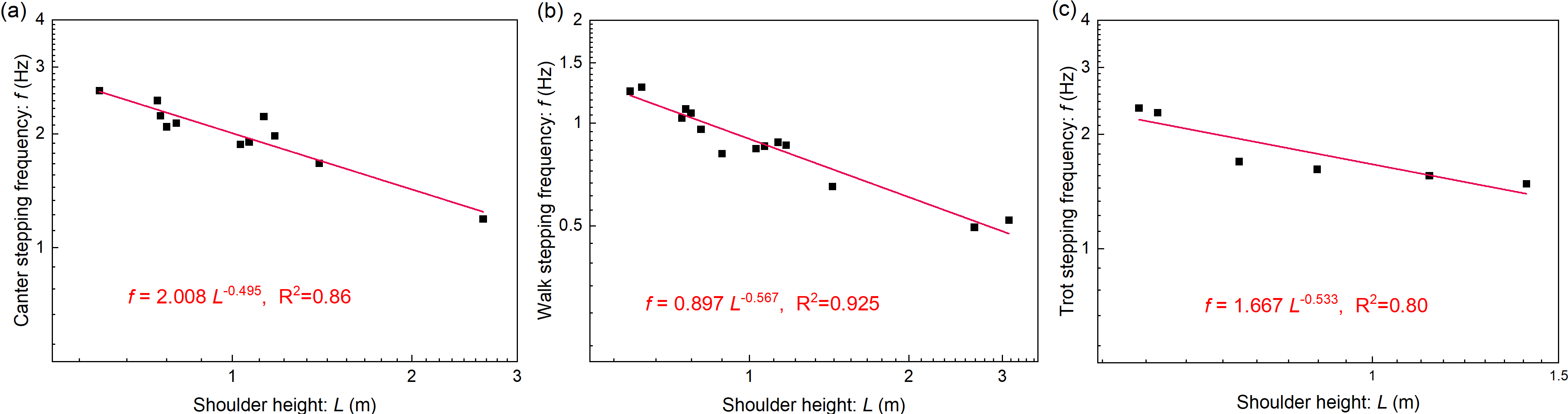}
\caption{Data reported by Pennycuick  \cite{pennycuick1975running} for legged animals.}\label{s14}
\end{figure}

We collect the data measured by Pennycuick \cite{pennycuick1975running} 
about the correlation between shoulder length and stepping frequency of mammals including {\it Thomson's gazelle: Gazella thomsonii, Warthog: Phacochoerus aethiopicus, Gnu: Connochaetes taurinus - calf,  Spotted hyaena: Crocuta crocuta, Grant's gazelle: Gazella granti, Impala: Aepyceros melampus, Lion: Panthera leo, Kongoni: Alcelaphus buselaphus, Topi: Damaliscus korrigum, Zebra: Equus burchelli,  Gnu: Connochaetes taurinus - adult, Black rhinoceros: Diceros bicornis, Giraffe: Giraffa camelopardalis,  Elephant: Loxodonta africana, and  Buffalo: Syncerus caffer.} See Figure \ref{s14}.

\vspace{2mm}

We collect the data reported by Gatesy and Biewener\cite{gatesy1991bipedal} about the stride frequency vs.~body mass and speed vs.~body mass in {\it Human.} 

\vspace{2mm}
The data collected by Blickhan and Full \cite{blickhan1993similarity} concerns body mass, stride frequency, and speed in various mammal species including {\it Dog: Canis familiaris, Ram: Ovis musimon, Human: Homo sapiens, Kangaroo-rat: Dipodomus merriami, Kangaroo-rat: Dipodomus spectabilis, {\rm and} Kangaroo: Megaleia rufa}.

\vspace{2mm}

We collect the data measured and summarized by Norberg and Rayner \cite{norberg2012scaling,norberg1987ecological} on the wing-beat frequency with body mass in bats including {\it Rousettus aegyptiacus, Eidolon helvum, Pteropus samoensis, Pteropus tonganus, Nyctimene robinssoni, Syconycteris australis, Taphozous australis, Lavia frons, Rhinolophus fumigatus, R. megaphyllus, Noctilio leporinus, Pteronotus parnelli, Mormoops blainvilli, Artibeus jamaicensis, Brachyphylla cavernarum, Erophylla bombifrons, Monophyllus redmani, Plecotus auritus, Nyctalus noctula, Miniopterus australis, Nyctophilus bifax, Myotis daubentoni, Pipistrellus pygmaeus, Eptesicus nilssoni, Tadarida brasiliensis, Tadarida pumila, Otomops martiensseni, Pteropus poliocephalus, Pteropus scapulatus, Taphozous hilli, Saccolaimus flaviventris, Craseonycteris thonglongyai, Macroderma gigas, Hipposideros ater, Rhinonycteris aurantius, Mormopterus planiceps, Chalinolobus gouldii, Chalinolobus morio, Chalinolobus nigrogriseus, Eptesicus serotinus, Myotis siliogorensis, Myotis dasycneme, Miniopterus schreibersi, Nyctalus noctula, Nyctophilus arnhemensis, Nyctophilus geoffroyi, Nyctophilus gouldii, Nyctophilus timoriensis, Scotorepens balstoni, Scotorepens greyi, Vespadelus (Pipistrellus) finlaysoni, Vespadelus regulus, Tadarida australis, Pteropus alecto, Pteropus poliocephalus, Hypsignathus monstrosus, Rhinolophus ferrumequinum, Rhinolophus hipposideridae, Glossophaga soricina, Carollia perspicillata, Desmodus rotundus, Myotis lucifugus, Myotis mystacinus, Myotis nattereri, Myotis sodalis, Eptesicus fuscus, {\rm and} Eptesicus serotinus.}

\vspace{2mm}
We summarize the data collected by Rodr\'iguez et al.~\cite{sanchez2023scaling} about the tail beat frequency and swimming speed vs.~body length of swimming mammals including {\it  Stenella frontalis, Lagenorhynchus obliquidens, Tursiops truncatus, Delphinapterus leucas, Pseudorca crassidens, Orcinus orca, Physeter macrocephalus,
Balaenoptera bonaerensis,
Megaptera novaeangliae,
Balaenoptera physalus,
Balaenoptera borealis,
Balaenoptera edeni,
Balaenoptera musculus,
Trichechus manatus latirostris,
Balaenoptera physalus,
Phoca hispida, {\rm and}
Phoca groenlandica.}

\vspace{3mm}
\noindent{\bf Crustacean}

\begin{figure}[H]
\centering
\includegraphics[width=0.6\textwidth]{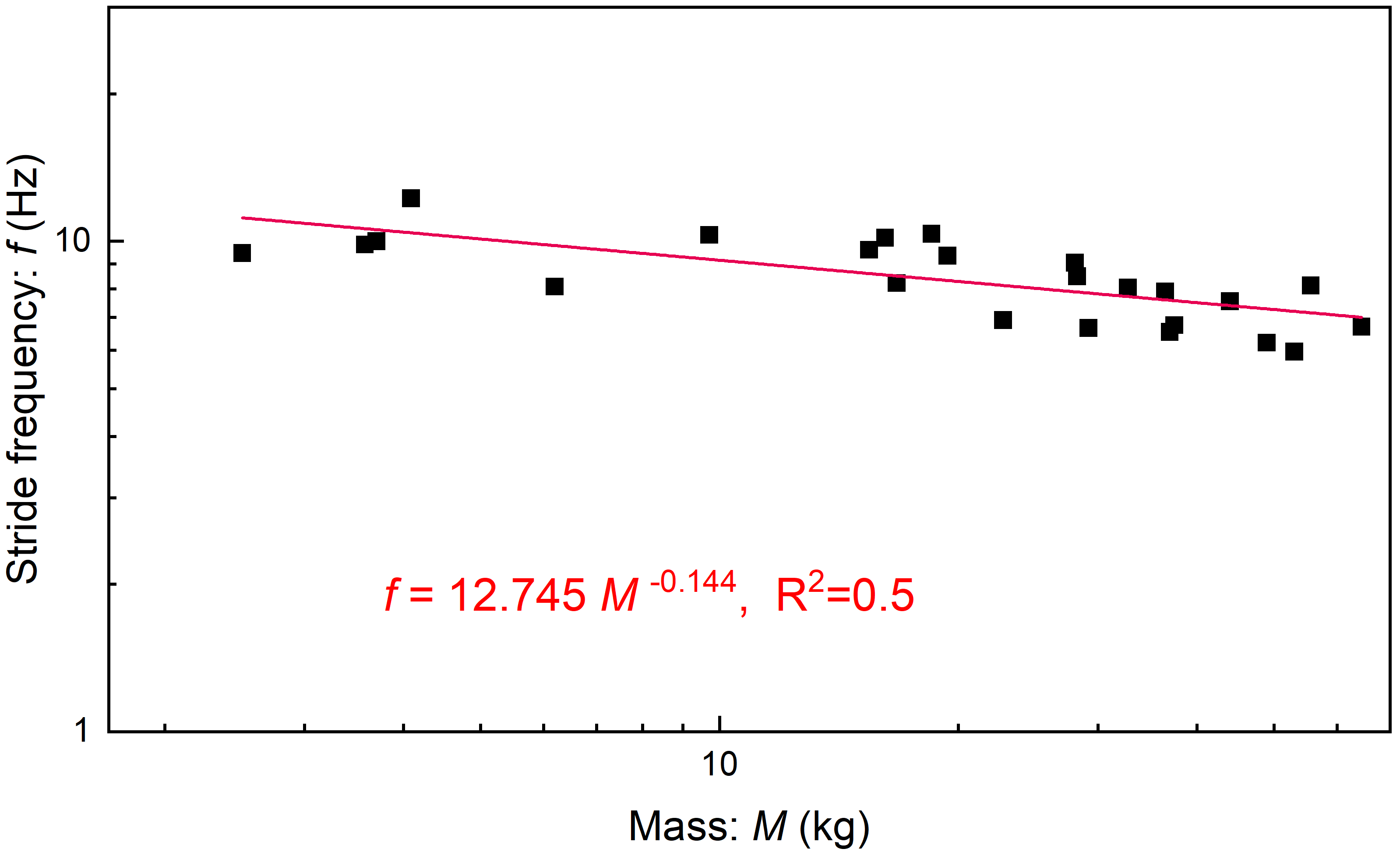}
\caption{Data reported by Whittemore et al. \cite{whittemore2015stride} and  Blickhan and Full \cite{blickhan1993similarity} for crabs.}\label{s15}
\end{figure}

\vspace{2mm}
The data on crabs measured by Whittemore et al. \cite{whittemore2015stride} and collected by Blickhan and Full \cite{blickhan1993similarity} is 
about stride frequency as a function of body mass in Ghost crabs: {\it Ocypode quadrata}. See Figure \ref{s15}.

We collect the data of shrimp measured by Arnott et al.~\cite{arnott1998tail} about the swimming speed vs.~body length and the tail-flip angular velocity vs.~body length in which the angular velocity is converted to frequency by dividing $2\pi$.  The species of the measured shrimp are {\it Brown shrimp Crangon crangon}.

\vspace{3mm}
\noindent{\bf Insects}
\vspace{2mm}

\begin{figure}[H]
\centering
\includegraphics[width=0.9\textwidth]{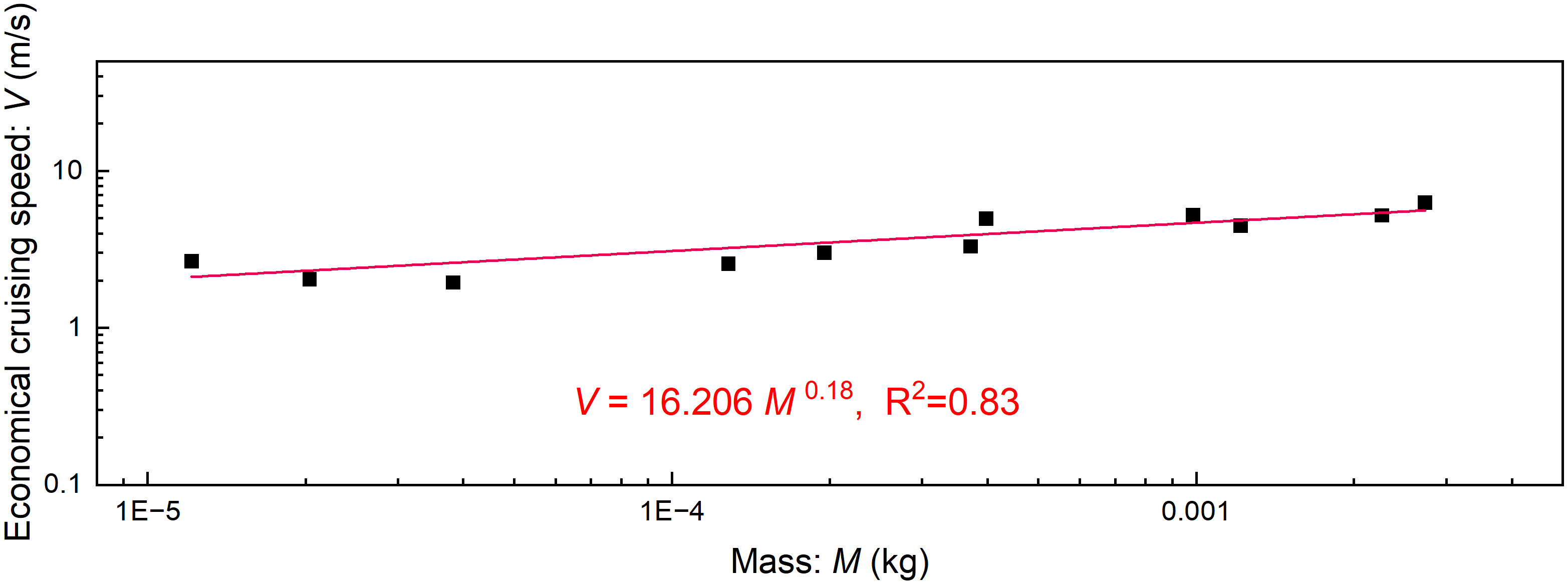}
\caption{Data reported by Greenewalt\cite{greenewalt1962dimensional} and Pennycuick \cite{pennycuick1969mechanics} for insects.}\label{s16}
\end{figure}

We summarize the data measured by Greenewalt \cite{greenewalt1962dimensional} and collected by Pennycuick \cite{pennycuick1969mechanics} is summarized in our Figure 1(C): the cruising speed vs.~body mass of flying animals including {\it Lucanus cervus, Sphinx convolvuli, Danaus plexippus, Melollontha vugaris, Bombus terrestris, Papilio machaon, Vanessa io, Pieris brassicae, Calliphora vomitoria, Apis mmellitica, Pyrrhosoma minimum, Lschnura elegant, Musca domestica}. The plot in \cite{pennycuick1969mechanics} shows the range of cruising speeds of different species. We extracted the middle point for each speed range. See Figure \ref{s16}.

\vspace{2mm}
We summarize the data collected by Blickhan and Full \cite{blickhan1993similarity} about the body mass, stride frequency, and speed in {\it Cockroach: Blaberus discoidalis}.

\begin{figure}[H]
\centering
\includegraphics[width=0.9\textwidth]{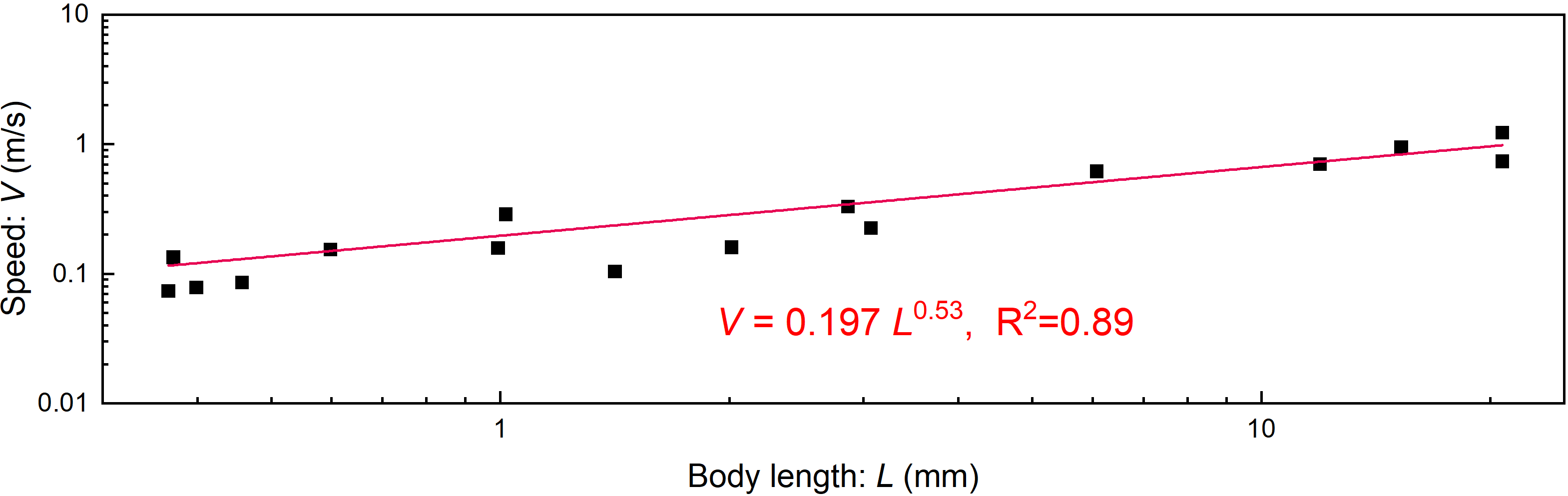}
\caption{Data reported by Farisenkov et al.\cite{farisenkov2020extraordinary} for insects.}\label{s17}
\end{figure}

\vspace{2mm}
We summarize the data measured and collected by Farisenkov et al.~\cite{farisenkov2020extraordinary} about the speed as a function of body length of beetles including {\it Acrotrichis grandicollis, A. sericans, Mikado sp., Nanosella sp., Nephanes titan, Paratuposa placentis, Ptenidium pusillum, Atheta sp., Dinaraea sp., Gyrophaena sp.~1, Gyrophaena sp.~2, Lordithon lunulatus, Philonthus sp., Nicrophorus investigator, Nicrophorus vespillo, {\rm and } Oiceoptoma thoracicum.} See Figure \ref{s17}.

\begin{figure}[H]
\centering
\includegraphics[width=\textwidth]{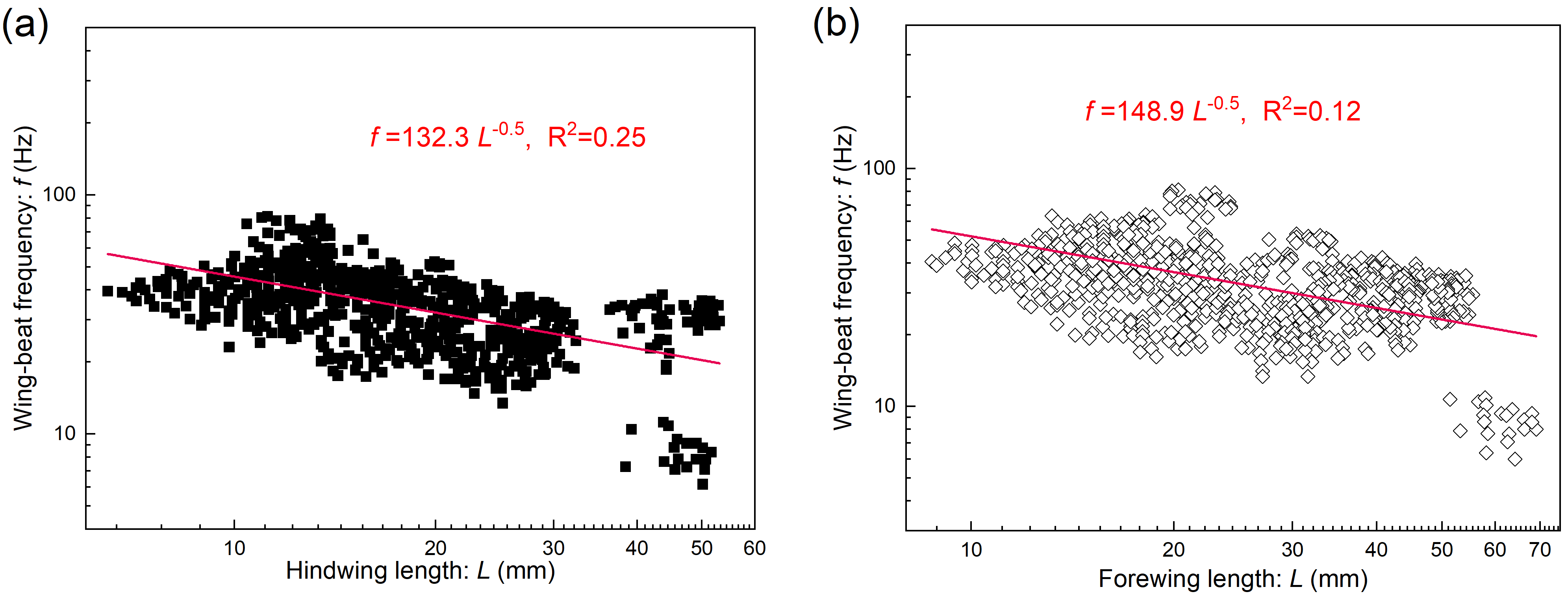}
\caption{Data reported by Yu et al.\cite{yu2022characterization} for insects.}\label{s18}
\end{figure}

We collect the data measured by Yu et al.~\cite{yu2022characterization} about wind-beat frequency and body mass of insects including 77 species in 14 families and 3 orders. The species are {\it  Mythimna separata, Agrotis ipsilon, Agrotis segetum, Argyrogramma agnata, Proxenus lepigone, Melicleptria scutosa, Enmonodia vespertili, Speiredonia retorta, Thyas juno, Chrysoruthrum amata, Trachea atriplicis, Hypocala subsatura, Prarlleila stuposa, Abrostola triplasia, Argyrogramma albostriata, Mythimna rufipennis, Polia illoba, Helicoverpa armigera, Heliothis dipsacea, Oraesia excavata, Spodoptera exigua, Spodoptera litura, Athetis dissimilis, Anomis mesogona, Agrotis tokionis, Macdunnoughia crassisigna, Eligma narcissus, Scotogramma trifolii, Leucania velutina, Brevopecten consanguis, Sidemia spilogramma, Acronicta major, Acronicta hercules, Mocis annetta, Psilogramma increta, Agrius convolvuli, Parum colligata, Pergesa elpenorlewisi, Clanis undulosa, Acosmeryx naga, Ampelophaga rubiginosa, Teretra japonica, Macroglossum stellatarum, Callambulyx tatarinovi,  Clanis bilineata, Palpita nigropunctalis, Maruca testulalis, Proteuclasta stotzneri, Diaphania indica, Mecyna gracilis, Botyodes diniasalis, Haritalodes derogata, Conogethes punctiferalis, Hymenia recurvalis, Diaphania quadrimaculalis, Theophila mandarina, Philosamia cynthia, Culcula panterinaria,
Calospilos suspecta, Amraica superans, Thosea sinensis, Latoia consocia, Cnidocampa flavescens, Hyphantria cunea, Spilarctia subcarnea, Rhyparioides amurensis, Chionarctia nivea M\'en\'etri\`es, Stilpnotia candida, Polygonia c-aureum, Vanessa cardui, Phalera grotei, Phalera takasagoensis, Pterostoma sinicum, Pantala flavescens, Anax parthenope julius, Chrysoperla sinica, Chrysopa phyllochroma, {\rm and} Chrysopa pallens}. See Figure \ref{s18}.

\begin{figure}[H]
\centering
\includegraphics[width=0.9\textwidth]{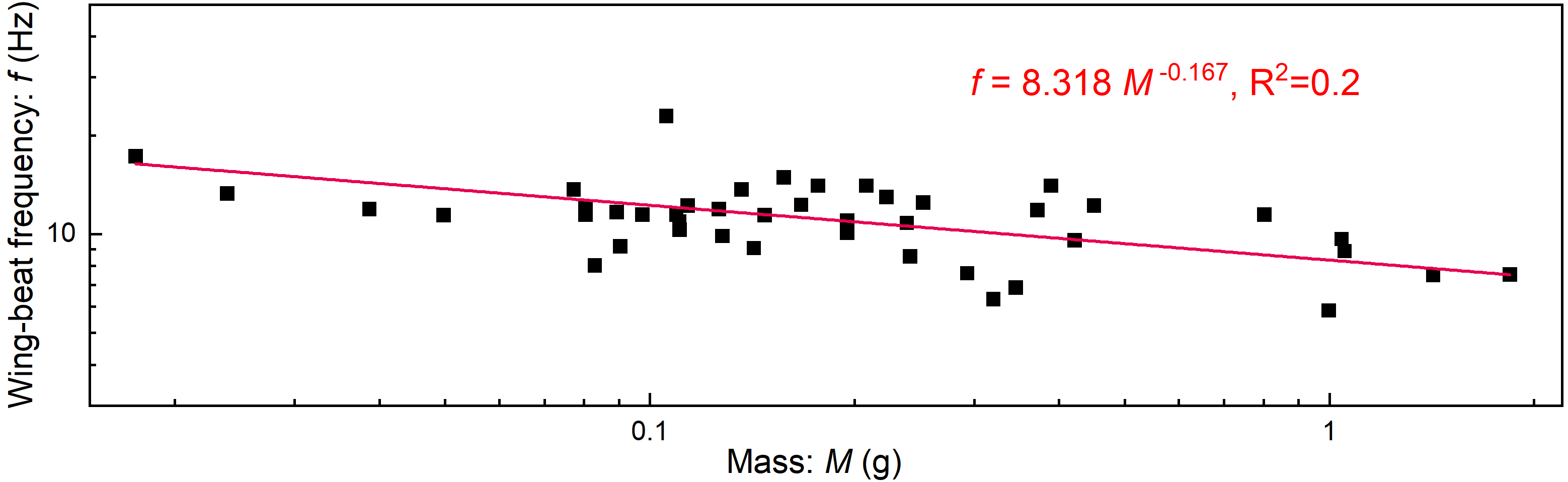}
\caption{Data reported by Dudley \cite{dudley1990biomechanics} for butterfies.}\label{s19}
\end{figure}

\vspace{2mm}
We collect the data measured by Dudley \cite{dudley1990biomechanics} about wing-beat frequency and body mass of 44 butterflies including species of {\it Battus polydamas (Linnaeus), Papilio thoas Rothschild and Jordan, Parides childrenae (Gray), Aphrissa boisduvalii (Felder), Itaballia demophile Joicey and Talbot, Archaeoprepona demophon Fruhstorfer, Myscelia cyaniris (Doubleday), Pyrrhogyra naerea Godman and Salvin, Siproeta stelenes (Fruhstorfer), Dryas iulia (Fabricius), Janatella leucodesma (Felder and Felder), Morpho amathonte Deyrolle, Morpho peleides Butler, Caligo illioneus Butler, {\rm and} Pierella luna (Fabricius).}See Figure \ref{s19}.

\begin{figure}
\centering
\includegraphics[width=0.9\textwidth]{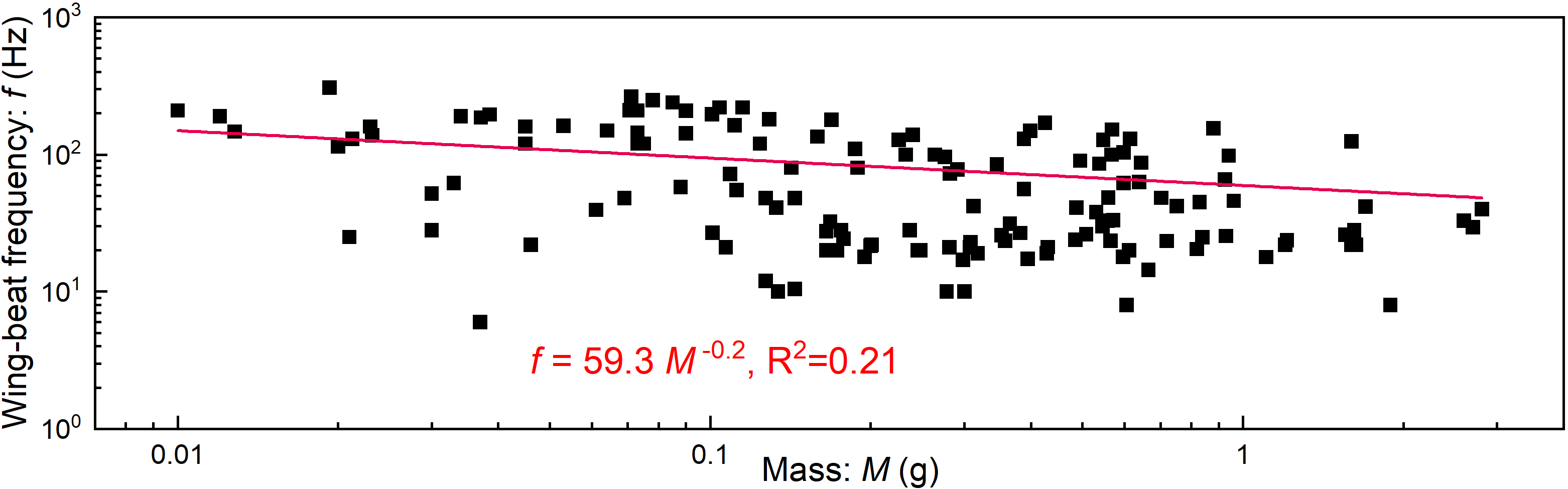}
\caption{Data reported by Byrne \cite{byrne1988relationship} for homopterous insects.}\label{s20}
\end{figure}

\vspace{2mm}
We collect the data measured and collected by Byrne \cite{byrne1988relationship} about relations between wingbeat frequency and body mass in homopterous insects including 154 species, which are {\it Bemisia tabaci, Trialeurodes vaporariorum, Dialeurodes citri, Trialeurodes abutilonea, Aleurothrixus floccosus, Aphis gossypii, Mysus persicae, Aphis fabae, Aphis nerii, Acyrthosiphon kondoi, Trichocera sp., Aedes aegypti, Drosophila viritis, Culicidae sp., Theobaldia annulata, Fannia scalaris, Musca domestica, Plalychirus peltalus, Sphaerophoria scripla, Syrphus grossulariae, Venilia macularia, Apis sp., Syrphus corollae, Syrphus nitens, Calliphora erythrocephala, Syrphus balteatus, Panorpa communis, Tipula sp., Ophion luteus, Catabomba pirastri, Pieris napi, Syrphus ribesii, Syrphus vitripennis, Ammophila sabulosa, Sarcophaga carnaria, Coenonympha pamphilus, Calliphora erythrocephala, Perithemis tenera, Chrysotoxum vernale, Tipula gigantea, Eristalis arbustorum, Euglossa saphirina, Chrysotoxum arcuatum, Eristalis tenax, Volucella pellucens, Chrysotoxum bicinctum, Apis mellifera, Malacosoma americanum, Euglossa mandibularis, Vespa vulgaris, Lymantria dispar, Euglossa dissimula, Rhodocera rhamni, Telephorus fuscus, Eristalis tenax, Poecilocampa populi, Polistes gallicus, Volucella plumata, Pieris brassicae, Zygoena filipendulae, Eristalis tenax, Agrostis exclamationis, Vanessa atalanta, Cerambycidae sp., Pieris brassicae, Plusia gamma, Bombus hortorum, Chelonia villica, Tetragoneuria cynosura, Hylesia spp., Euglossa imperialis, Vanesa cardui, Erythemis simplicicollis, Pachydiplax longipennis, Vespa germanica, Macroglossa bombyliformis, Vanessa io, Hyperchirica nausica, Notodonta dictaea, Bombus muscorum, Dasyramphis atra, Dasichyra pudibunda, Vespa germanica, Libellula depressa, Orthetrum coenilescens, Chrysocoris purpureus, Tabanus bovinus, Argynnis pandora, Sympetrum meridionale, Macroglossa stelatorum, Amphimallon solstitialis, Automeris jacunda, Papilio podalirius, Leptetrum quadrimaculatum, Pantala flavescens, Ophiogomphus serpentinus, Libellula luctuosa, Macroglossa stellatarum, Somatochlora tenebrosa, Tramea carolina, Plathemis lydia, Tramea lacerata, Bombus terrestris, Enyo ocypete, Antomeris fieldi, Eulaema nigrita, Eufriesia pulchra, Automeris zugana, Nasiaeschna pentacantha, Triphoena pronuba, Adeloneivaia subungulata, Bombus lapidarius, Libellula pulchella, Aeschna mixta, Celonia aurata, Macromia georgina, Eulaema cingulata, Brachytron pratense, Xylophones libya, Automeris hamata, Vespa crabro, Bombus lapidarus, Manduca lefeburei, Bombyx rubi, Melolontha vulgaris, Vespa crabro, Philosamia cynthia, Aeschna rufescens, Xylocopa violacea, Perigonia lusca, Exaerete frontalis, Automeris belti, Pachygonia drucei, Automerina auletes, Cicada sp., Anax junius, Xylophones pluto, Adelotieivaia boisduvalii, Bombus terrestris, Tesseraloma javanico, Macromia taeniolata, Eulaema meriana, Melolontha vulgaris, Eacles imperialis, Anax formosus, Erinnyis ello, Periplanela americana, Acheronlia atropos, Bombus sp., Manduca corallina, Syssphinx molina, Madoryx oeclus, Saturnia pyri, Lucanus cervus, Manduca rustica, {\rm and} Oryba achemenides}. See Figure \ref{s20}.

\vspace{2mm}
We include the data of 30 individuals of eight insect species from five orders measured by Ha et al.~\cite{san2013relationship} about wing-beat frequency and body mass. The eight species are {\it Sympetrum flaveolum, Pieris rapae, Plusia gamma, Ochlodes, Xylocopa pubescens, Bombus, Tibicen linnei, {\rm and} Allomyrina dichotoma.}

\vspace{2mm}
We summarize the data collected by Tennekes \cite{tennekes2009simple} about the weight and cruising speed of insects including {\it midge, hover fly, gnat, house fly, crane fly, meat fly, honeybee, hornet, bumblebee, summer chafer, little stag beetle, cock chafer, dung beetle, stag beetle, damsel fly, scorpion fly, green-veined white, ant lion, cabbage white, green dragonfly, common swallowtail butterfly, eyed hawk moth, yellow-banded dragonfly, blue underwing moth, {\rm and} privet hawk moth}. The weight has been converted to body mass to integrate into Figure 1(B), also see Figure \ref{simple_science}.

\vspace{2mm}
We include the data of about 35
sphingids, 50 saturniids, and 20 other heterothermic moths, which is measured by Bartholomew and Casey \cite{bartholomew1978oxygen} about the body mass and wing-beat frequency. The species include {\it Manduca rustica, Manduca corallina, Protambulyx smmgil, Erinnyis oenotrus, Pachylia ficus, Madoryx oeclus, Perigonia lusca, Enyo ocypete, Oryba achemenides, Pachygonia drucei, Xylophanes libya, Automeris hamata, M. lefeburei, E. ello, M. oeclus, X. chiron, X. pluto, A. jacunda, A. belti, A. fieldi, A. zugana, Automerina auletes, Pseudoautomeris salmonea, Dirphea agirs, Hylesia praeda, H. sp., Hyperchirica nausica, Eacles imperiali, Sphingicampa quadrilineata, Syssphinx molina, Adeloneivaia subungulata, A. boisduvalii, Rhescyntis hippodamia, Titaea tamerlan, Rothschildea orizaba, Neocossus sp., Megalpyge sp., Pyralidae: Genus sp., Artace sp., {\rm and} Noctuidae: Genus sp}. 

\begin{figure}[H]
\centering
\includegraphics[width=0.9\textwidth]{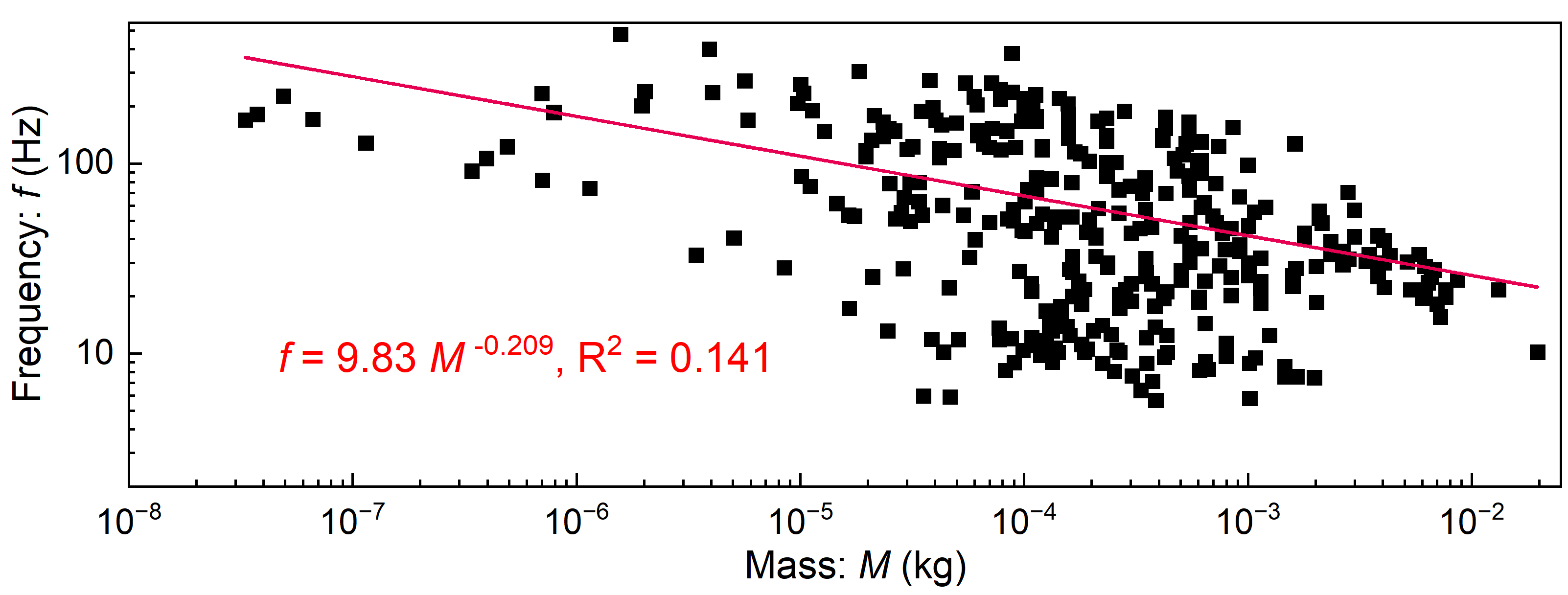}
\caption{Data collected by Dudley \cite{dudley2002biomechanics} on wind-beat frequency versus body mass for a wide variety of insects.}\label{s21}
\end{figure}

\vspace{2mm}
We collect the data summarized by Dudley \cite{dudley2002biomechanics} on the wing-beat frequency and mass of insects, including species of {\it Brattaria, Coleoptera, Diptera, Hemiptera, Heterocera, Homoptera, Hymenoptera, Mecoptera, Neuroptera, Odonata, Orthoptera, Rhopalocera} and {\it Trochilidae}. See Figure \ref{s21}.

\vspace{3mm}
\noindent{\bf Microorganisms}

\vspace{2mm}
We collect the data measured by Kamdar et al.~\cite{kamdar2023multiflagellarity} about speed as a function of body length in monotrichous bacteria including {\it Vibrio alginolyticus, Vibrio natriegens, {\rm and} Rhodobater sphaeroides}, where the speed has been normalized by the real speed of $V_0$ of the smallest observed bacteria.
\begin{figure}[H]
\centering
\includegraphics[width=0.9\textwidth]{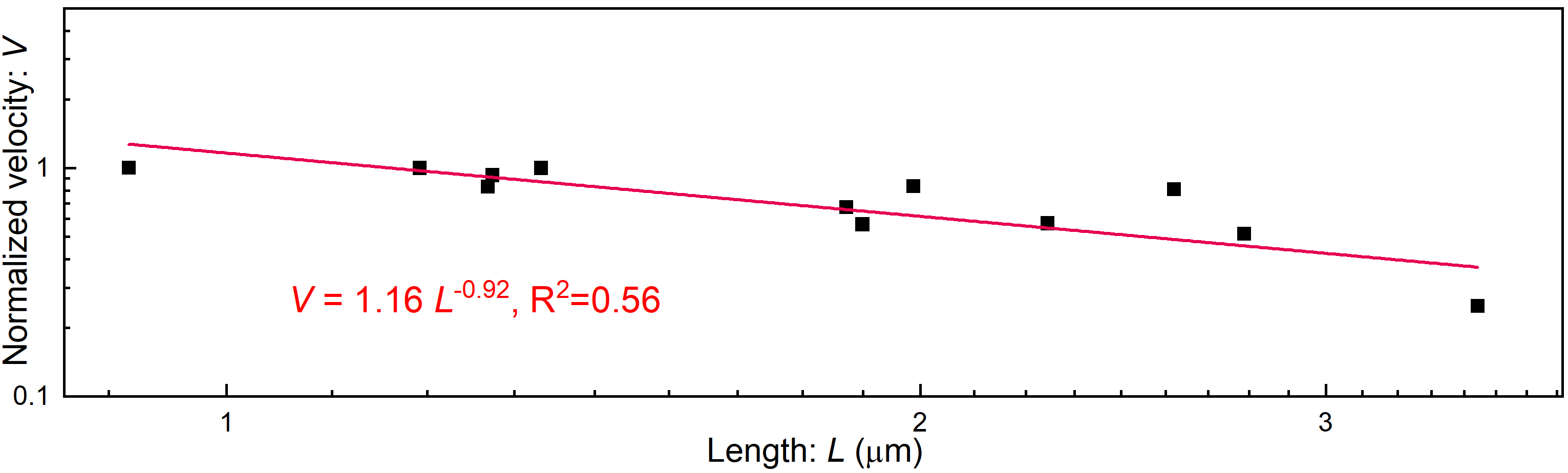}
\caption{Data reported by Kamdar et al.\cite{kamdar2023multiflagellarity} for monotrichous bacteria, where the normalized velocity is defined by $V={\rm real\,speed}/V_0$ with $V_0$ the velocity of the smallest observed bacteria.}
\end{figure}

\vspace{2mm}
We collect the data measured by Cramer et al.~\cite{cramer2021longer} about swimming speed vs.~the total length of sperm in the canary islands chiffchaff ({\it phylloscopus canariensis}) from EI Hierro, Gran Canaria, La Palma, and Tenerife.
The data measured by Helfenstein et al.~\cite{helfenstein2008between} about swimming speed vs.~the total length of sperm in sand martins {\it Riparia riparia} also shows the inversely proportional relationship.

\vspace{3mm}
\noindent{\bf Plants}

\begin{figure}
\centering
\includegraphics[width=0.9\textwidth]{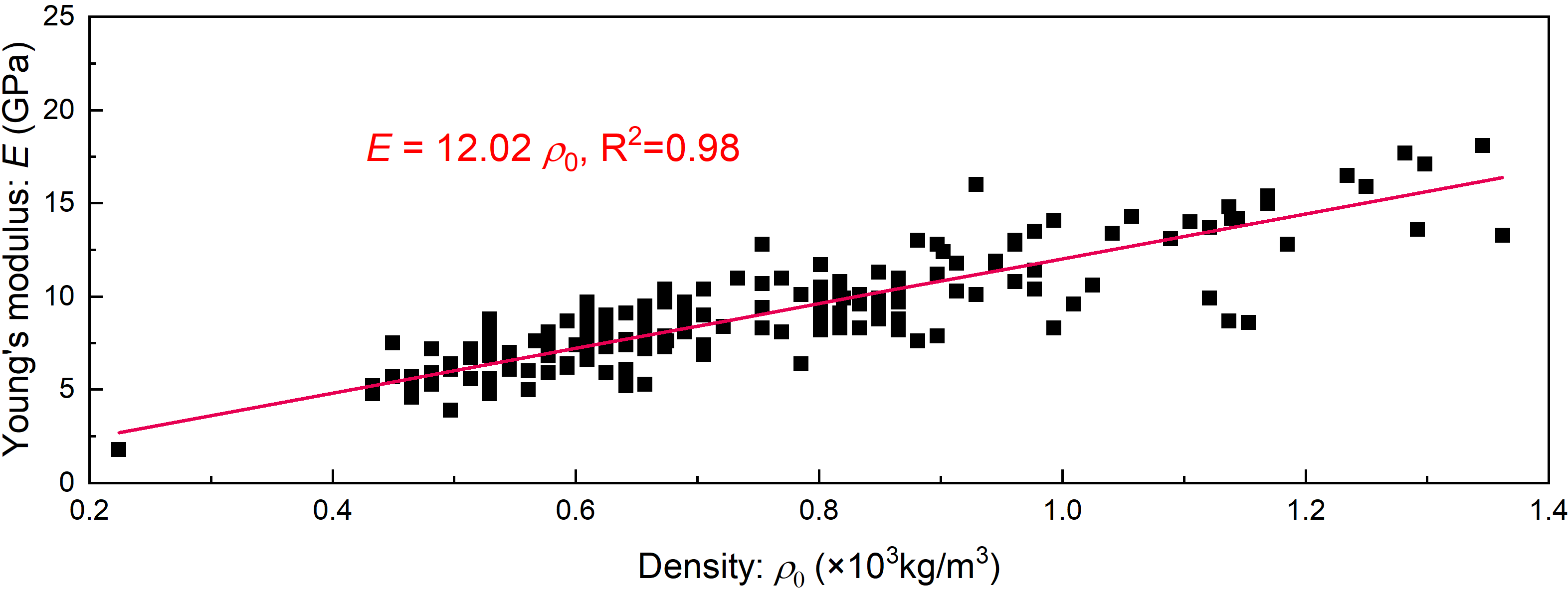}
\caption{Data reported by Niklas and Spatz\cite{niklas2010worldwide} about Young's modulus as a function of density for trees.}\label{s22}
\end{figure}

\vspace{2mm}
We summarize the data of 161 species of trees measured by Niklas and Spatz \cite{niklas2010worldwide} about Young's modulus as a function of density. The species are {\it Abies alba, Abies balsama, Abies grandis, Abies procera, Agathis vitiensis, Araucaria angustifolia, Chaemaecyparis lawsoniana, Larix decidua, Larix eurolepis, Larix kaempferi, Picea abies, Picea alba, Picea omorika, Picea sitchensis, Pinus brutia, Pinus caribaea, Pinus holfordiana, Pinus nigra, Pinus pinaster, Pinus ponderosa, Pinus radiata, Pinus strobus, Pinus sylvestris, Podocarpus   sp., Podocarpus guatemalensis, Pseudotsuga menziesii, Thuja heterophylla, Thuja plicata, Aesculus hippocastanum, Acacia mollissima, Acer psedudoplatanus, Afzelia quanzensis, Alnus glutinosa, Alstonia boonei, Anthocephalus chinensis, Aspidosperma sp., Autranella congolensis, Berlinia confusa, Betula sp., Brachstegia nigerica, Brachylaena hutchinsii, Byrsonima coriacea, Calophyllum brasiliense, Canarium schweinfurthii, Carpinus betulus, Cassispourea malasana, Castanea sativa, Cedrela odorata, Celtis   sp., Ceratopetalum apetalum, Chlorophora excelsa, Cordia millenii, Cullenia ceylanica, Cyanomeria alexandri, Cylicodiscus gabunensis, Dipterocarpus sp., Dipterocarpus acutangulus, Dipterocarpus caudiferus, Dipterocarpus zeylanicus, Drychalanops beccarii, Drychalanops keithii, Drychalanops lanceolata, Entandrophragma angolense, Entandrophragma cylindricum, Entandrophragma utile, Eperua sp., Erythrophteum   sp., Eucalyptus piluaris, Eucalyptus marginata, Eucalyptus microcorys, Eucalyptus paniculata, Eucalyptus versicolor, Eusideroxylon zwageri, Fagus sylvatica, Fraxinus excelsior, Gmelina arborea, Gonystylus macrophyllum, Gossweilerodendron balsamiferum, Guarca excelsa, Guarca thompsonii, Heritiera simplicifolia, Hevea brasiliensis, Hopea sengal, Khaya anthotheca, Khaya grandifl ora, Khaya ivorensis, Khaya nyascia, Koordersiodendron pinnatum, Loniciocarpus castillo, Lophira alata, Lovoa trichilioides, Maesopsis veminii, Mansonia altissima, Mora excelsa, Muragne sp., Nauclea diderrichii, Nectrandra sp., Newtonia buchaneni, Nothofagus sprocera, Ocotea rodiaei, Ocotea usambarensis, Octomeles sumatrana, Olea hochstetteri, Oxystigma oxyphyllum, Parashorea   sp., Parashorea malaanonan, Parashorea tomentelia, Peltogyne   sp., Pericopsis elata, Pipradeniostrum africanum, Platanus hybrida, Populus canadensis, Populus x canescens, Protium decendrum, Prunus avium, Pseudosindora palustris, Pterocarpus angolensis, Pterygota bequaertii, Pterygota macrocarpa, Quercus   sp., Quercus cerris, Quercus rubra, Ricinodendron rautanenii, Salix x alba, Salix alba var. coerulea, Salix fragilis, Sclerocarpa   sp., Scottellia coriacea, Shorea acuminatissima, Shorea dasphylla, Shorea faguetiana, Shorea gibbosa, Shorea guiso, Shorea hakeifolia, Shorea leptoclados, Shorea macrophylla, Shorea parvifloia, Shorea pauciflora, Shorea smithiana, Shorea superba, Shorea superba, Shorea waltonii, Staudtia stipitata, Sterculia oblonga, Sterculia rhinopetala, Swartzia leiocalycine, Symphonia globulifera, Syncarpia glomulifera, Tarrietia utilis, Tectona grandis, Terminalia amazonica, Tieghemelia heckerii, Tilia vulgaris , Triplochiton scleroxylon, Ulmus glabra, Ulmus hollandica, Ulmus procera, Virola koschnyi, Vochvsia   sp., {\rm and} Vochysia hondurensis}. See Figure \ref{s22}.

\begin{figure}[H]
\centering
\includegraphics[width=0.9\textwidth]{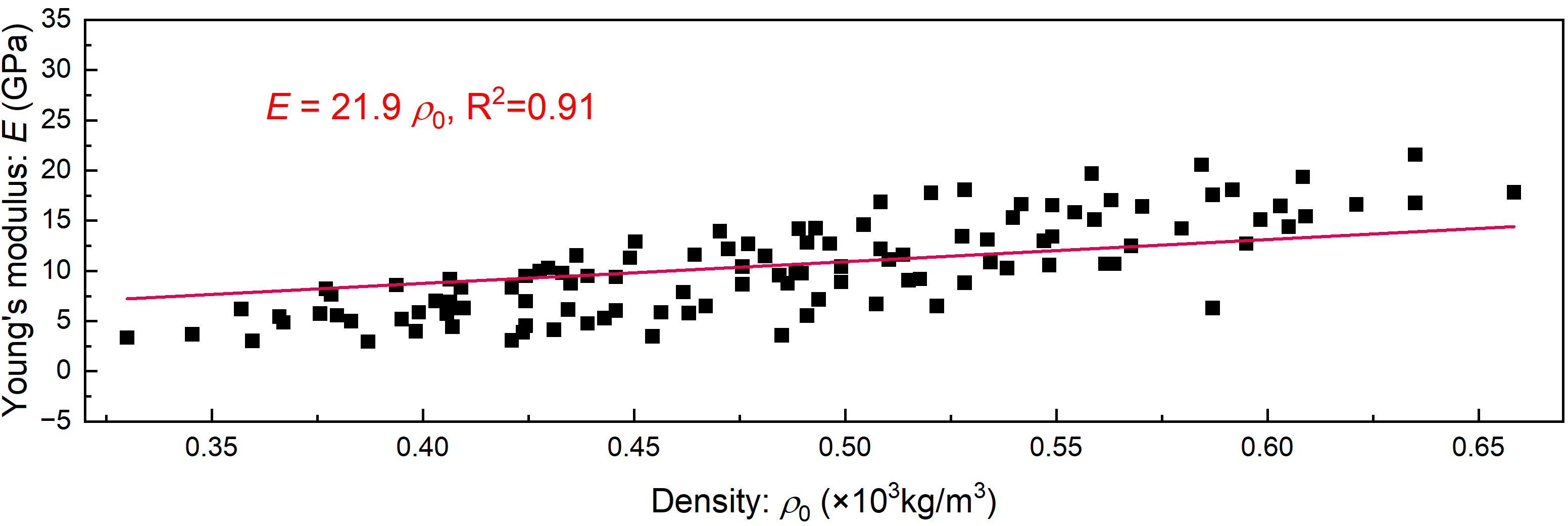}
\caption{Data reported by Watt et al.\cite{watt2010determining} of Young's modulus as a function of density for trees.}\label{s23}
\end{figure}

\vspace{2mm}
We include the radial measured data for 13-year-old radiata pine {\it Pinus radiata} clones by Watt et al.~\cite{watt2010determining} about the modulus of elasticity vs.~density. See Figure \ref{s23}.

\begin{figure}
\centering
\includegraphics[width=0.9\textwidth]{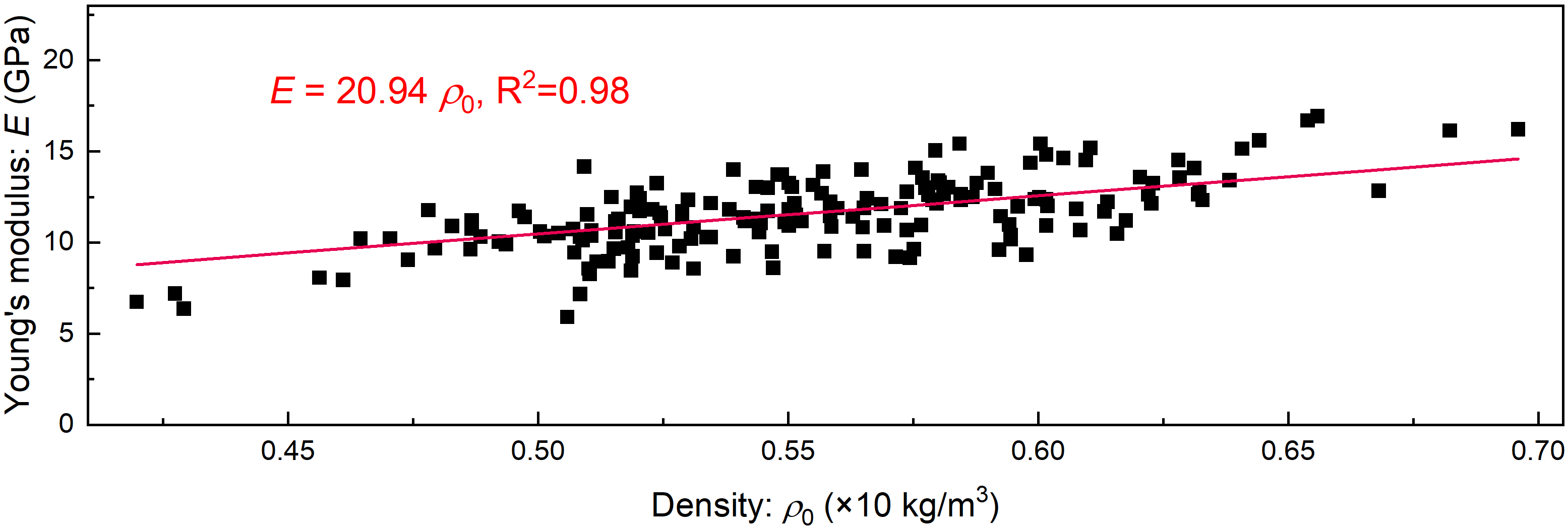}
\caption{Data reported in Lachenbruch et al.\cite{lachenbruch2010relationships} of Young's modulus as a function of density for trees.}\label{s24}
\end{figure}

\vspace{2mm}
We collect the data measured by Lachenbruch et al.~\cite{lachenbruch2010relationships} about modulus of elasticity vs.~density for Douglas-fir {\it Pseudotsuga menziesii (Mirb.) Franco}. The measurements include 1087 specimens in total in which 183 trees over 20 years old and 6 species are included. See Figure \ref{s24}.

\vspace{2mm}
We collect the data measured by Ngadianto et al.~\cite{ngadianto2020wood} about the dynamic modulus of elasticity vs.~density of stems of three tree species grown in Yogyakarta, Java Island, and
Indonesia, which include {\it Acacia mangium Willd., Maesopsis eminii Engl., {\rm and} Melia azedarach L.}

\vspace{2mm}
McMahon \cite{mcmahon1973size} mentioned that the density-specific modulus $E/\rho_0$ is nearly constant in the green wood, which is concluded from the data measured by Garratt \cite{garratt,wangaard1950mechanical} and McElhanney and Perry \cite{McElhanney}.

\subsubsection*{Running speed of King Kong 2005}

The speed of King Kong varies depending on the source. In this paper we base our measurement on direct observations from the 2005 movie. The specific video clip referenced covers the period from 02:01:34 to 02:01:41 when King Kong is running through the forest. Assume the average width of the reference big trees is 15 meters, and given that King Kong can cross 2 to 3 trees in one second, his speed is approximately 40 $m/s$.
\bibliographystyle{unsrt}
\bibliography{scaling0}

\end{document}